# Stimulated and Spontaneous Emission of Radiation in a Single Mode for N-TLMs

Michael T. Tavis and Frederick W. Cummings[*]

## Abstract

The general expressions for the time development of the ensemble averages of $E^-$ and $E^-E^+$ are found for N two level molecules (TLMs) interacting with a single mode electromagnetic field. The TLMs are assumed to be at equivalent mode positions in the field. In the main body of the paper, the general results are then simplified by assuming that all TLMs are in the up state at time 0. Examples of spontaneous emission behavior including both Super Radiant behavior and emission suppression are shown in Appendix E while stimulated emission with all TLMs initially in the lower state is examined in Appendix F. Both resonant and non resonant cases are considered in the main body. Analytic results are presented for low numbers of TLMs. In the appendices, it will be shown that for the mean photon number large in comparison to the number of TLMs, that the Average Field Approximation (AFA) is a useful representation with much simpler analytic form. The validity of this approximation will also be examined. In the main paper, a comparison between the exact results and results using the AFA will be presented for a small number of TLMs and a large average photon number. It will be found that the results are in general remarkably similar but for non resonance differ somewhat at a finer level of detail. Of interest is that the ensemble averages for N TLMs is similar to the ensemble average for a single TLM multiplied by the number of TLMs which is somewhat disappointing but is believed to be the consequence of assuming that all TLMs are at equivalent mode positions and large mean photon number. Some results for smaller photon number will also be presented

## Introduction

In 1965, Cummings[1], published the classical paper on the time development of the ensemble averages of $E^-$ and $E^-E^+$ for photon number densities representing the Glauber State[2] and thermal state but only for the resonant case. His presentation did not include large values of time. Eberly[3] showed, in a refereed paper, the long term development for the ensemble average of $E^-E^+$ which showed the periodic Collapse and Revival of oscillations in that average. Finally, in 1987, a demonstration of

---

[*] Michael Tavis (retired), Frederick W. Cummings (Professor Emeritus, University of California, Riverside)
[1] F. W. Cummings, Stimulated Emission of Radiation in a Single Mode, Physical Review, 140, #4A, p1051, 1965.
[2] Roy J. Glauber, Coherent and Incoherent States of the Radiation Field, Physical Review, Vol. 131, #6, Sept. 1963, p. 2766.
[3] J. H. Eberly, N. B. Narozhny, and J.J. Sanchez-Mondrgon, Periodic Spontaneous Collapse and Revival in a Simple Quantum Model, Phys. Rev. Lett. 44, 1323-1326 (1980)

quantum collapse and revival was observed in a one-atom maser[4]. Since that time numerous papers including several Wikipedia articles have been published. More recently, Tavis and Cummings presented results for Stimulated Emission of Radiation in a Single Mode for both Resonant and Non-resonance for various Initial Photon Density Distributions[5]

In the meantime, Tavis[6] provided the eigenvalues and eigenvectors for N TLMs interacting with a quantized radiation field. Buried in Appendix F of his thesis, he presented the general formulation for time development for the ensemble averages of $E^-$ and $E^-E^+$ for any number of TLMs.

In the following, we will review the development of this general solution, simplified for the case of all TLMs initially in the excited state and display analytic solutions for small numbers of TLMs for both resonant and non resonant cases. Examination of these simple cases will suggest an approximate solution which is developed in Appendices A and B. Numerical results for both exact and approximations are provided in the discussion section. Additional analytic results for 3 and 4 TLMs for the resonant case are presented in Appendix C while a discussion on the difficulty of presenting the non-resonant case is given in Appendix D. Spontaneous Emission for all TLMs in the up state, for all except 1 or 2 TLMs in the down state (suppression of emission is seen) and half the TLMs in the up state (super radiant behavior is seen) are given in Appendix E. Finally, Spontaneous Absorption for all TLMs in the down state is provided in Appendix F.

## Algebraic Background

### ENSEMBLE AVERAGES FOR $E^-$ AND $E^-E^+$

The ensemble average of the field operators $E^-(t)$ and $E^-E^+(t)$ may be found in the usual way[7]

$$\langle E^-(t)\rangle = -\left(\frac{\gamma}{\mu}\right)\sum_{n=0}^{\infty}(n+1)^{1/2}\langle n|\rho_f(t)|n+1\rangle, \tag{1a}$$

and

---

$$\langle E^- E^+(t) \rangle = \left|\frac{\gamma}{\mu}\right|^2 \sum_{n=0}^{\infty} n \langle n|\rho_f(t)|n \rangle, \tag{1b}$$

where only one mode of the field is excited, $\gamma$ is the complex coupling constant, and $\mu$ the dipole moment of the TLM with which the field is interacting. The element of the field density matrix is given by the trace over the TLM states

$$\langle n|\rho_f(t)|n'\rangle = \sum_{r,m} P(r) \langle n|\langle r,m|\rho(t)|r,m\rangle|n'\rangle \tag{2}$$

where

$$P(r) = \frac{N!(2r+1)}{\left(\frac{N}{2}+r+1\right)!\left(\frac{N}{2}-r\right)!} \tag{3}$$

and $\rho(t)$ is given by a unitary transformation of the density operator at time $t_o = 0$ where it is assumed that the N-TLMs and radiation field are not interacting! Therefore

$$\rho(t) = U(t)\rho(0)U^{-1}(t), \tag{4}$$

$$U(t) = e^{iHt}, \tag{5}$$

and where H is given[6] by

$$H = (a^\dagger a + R_3) + a^\dagger a \frac{\omega - \Omega}{\Omega} - \kappa a R_+ - \kappa^* a^\dagger R_- \tag{6}$$

Since the system is non-interacting at time zero, the density operator is a direct product of the field part and N-TLM part of the system.

$$\rho(0) = \rho_{TLM} \otimes \rho_f. \tag{7}$$

Equation (2) can be expanded directly in terms of the orthonormal eigenvectors of the Hamiltonian above, namely the |r,c,j> states which are expressed by[6],

$$|r,c,j> = \sum_{n=Max[0,c-r]}^{c+r} A_n^{r,c,j} |n>|r,c-n>. \tag{8}$$

In expression (8), r and c are good quantum numbers where $r \leq \frac{N}{2}$ and c=n+m with $-r \leq m \leq r$.

$$\langle n|\rho_f(t)|n'\rangle = \sum_{r,m} P(r) \sum_{r',c',j'r'',c'',j''} \langle n|\langle r,m|e^{-iHt}|r',c',j'\rangle \\ \times \langle r',c',j'|\rho(0)|r'',c'',j''\rangle \langle r'',c'',j''|e^{iHt}|r,m\rangle|n'\rangle. \tag{9}$$

As an alternative expansion of eq. (2) we can follow the development used by Cummings (1) for his similar derivation for the single TLM case, namely

$$\langle n|\rho_f(t)|n'\rangle = \sum_{r,m} P(r) \langle n|\langle r,m|\rho(t)|r,m\rangle|n'\rangle \\ = \sum_{r,m} \sum_{k=0}^{\infty} \sum_{r',p} \sum_{k'=0}^{\infty} \sum_{r'',p'} P(r)\langle n|\langle r,m|U(t,0)|r',p\rangle|k\rangle \\ \times \langle k|\langle r',p|\rho_{TLM}(0) \otimes \rho_f(0)|r'',p'\rangle|k'\rangle \langle k'|\langle r'',p'|U^{-1}(t,0)|r,m\rangle|n'\rangle. \tag{10}$$

Either expansion can be used as the results will be the same. The first expansion was used in the thesis listed in Reference (6) appendix F. Here the second will be used to be consistent with Reference (1). Consider the factors in eq. (10) separately

$$\langle n|\langle r,m|U(t,0)|r',p\rangle|k\rangle \\ \langle k|\langle r',p|\rho_{TLM}(0) \otimes \rho_f(0)|r'',p'\rangle|k'\rangle \\ \langle k'|\langle r'',p'|U^{-1}(t,0)|r,m\rangle|n'\rangle \tag{11}$$

Equation (8), can be inverted to express $|n>|r,c-n>$ in terms of $|r,c,j>$, namely

$$|n>|r,c-n> = \sum_{j=0}^{Min[2r,c+r]} (A^*)_n^{r,c,j} |r,c,j> \tag{12}$$

Using

$$U(t,0)|n>|r,c-n> = \sum_{j=0}^{Min[2r,c+r]} (A^*)_n^{r,c,j} e^{-i\lambda_{r,c,j}t} |r,c,j>, \tag{13}$$

the first term in eq. (11) can be written as

$$\langle n|\langle r,m|U(t,0)|r',p\rangle|k\rangle = \sum_{j=0}^{Min[2r,n+m+r]} (A^*)_k^{r,n+m,j} A_n^{r,n+m,j} e^{-i\lambda_{r,n+m,j}t} \delta_{n+m,k+p} \delta_{r,r'}. \tag{14}$$

In a similar fashion, $\langle k'|\langle r'',p'|U^{-1}(t,0)|r,m\rangle|n'\rangle$ can be express as

$$\langle k'|\langle r'',p'|U^{-1}(t,0)|r,m\rangle|n'\rangle = \sum_{j'=0}^{Min[2r,n'+m+r]} (A^*)_{n'}^{r,n'+m,j'} A_{k'}^{r,n'+m,j} e^{i\lambda_{r,n'+m,j'}t} \delta_{n'+m,k'+p'} \delta_{r,r''}. \tag{15}$$

The final term in eq. 13 is given by

$$\langle k|\langle r',p|\rho_{TLM}(0) \otimes \rho_f(0)|r'',p'\rangle|k'\rangle = \langle k|\rho_f(0)|k'\rangle \langle r',p|\rho_{TLM}(0)|r'',p'\rangle. \tag{16}$$

Equation (10) for the density matrix for the field at time t, is given by

$$\begin{aligned}
\langle n|\rho_f(t)|n\rangle &= \sum_{r=0,\frac{1}{2}}^{\frac{N}{2}} P(r) \sum_{m=-r}^{r} \sum_{k=0}^{\infty} \sum_{p=-r}^{r} \sum_{k'=0}^{\infty} \sum_{p'=-r}^{r} \sum_{j=0}^{Min[2r,n+m+r]} \sum_{j'=0}^{Min[2r,n'+m+r]} (A^*)_{k}^{r,n+m,j} \\
&\times A_n^{r,n+m,j} e^{-i\lambda_{r,n+m,j}t} (A^*)_{n'}^{r,n'+m,j'} A_{k'}^{r,n'+m,j} e^{i\lambda_{r,n'+m,j'}t} \langle k|\rho_f(0)|k'\rangle \\
&\times \langle r,p|\rho_{TLM}(0)|r,p'\rangle \delta_{n+m,k+p} \delta_{n'+m,k'+p'}
\end{aligned} \tag{17}$$

With this expression, equations (1a) and (1b) can be written as[8]

$$\begin{aligned}
\langle E^-(t)\rangle &= -\left(\frac{\gamma}{\mu}\right) \sum_{n=0}^{\infty} (n+1)^{1/2} \langle n|\rho_f(t)|n+1\rangle \\
&= -\left(\frac{\gamma}{\mu}\right) \sum_{r=0,\frac{1}{2}}^{\frac{N}{2}} P(r) \sum_{m=-r}^{r} \sum_{c=m}^{\infty} (c-m+1)^{1/2} \sum_{p=-r}^{r} \sum_{j=0}^{Min[2r,c+r]} \sum_{p'=-r}^{r} \sum_{j'=0}^{Min[2r,c+r+1]} (A^*)_{c-p}^{r,c,j} \\
&\times A_{c-m}^{r,c,j} (A^*)_{c-m+1}^{r,c+1,j'} A_{c-p'+1}^{r,c+1,j'} e^{-i\lambda_{r,c,j}t} e^{i\lambda_{r,c+1,j'}t} \langle c-p|\rho_f(0)|c+1-p'\rangle \langle r,p|\rho_{TLM}(0)|r,p'\rangle, \\
\langle E^-E^+(t)\rangle &= \left|\frac{\gamma}{\mu}\right|^2 \sum_{n=0}^{\infty} n\langle n|\rho_f(t)|n\rangle \\
&= \left|\frac{\gamma}{\mu}\right|^2 \sum_{r=0,\frac{1}{2}}^{\frac{N}{2}} P(r) \sum_{m=-r}^{r} \sum_{c=m}^{\infty} (c-m) \sum_{p=-r}^{r} \sum_{j=0}^{Min[2r,c+r]} \sum_{p'=-r}^{r} \sum_{j'=0}^{Min[2r,c+r]} (A^*)_{c-p}^{r,c,j} \\
&\times A_{c-m}^{r,c,j} (A^*)_{c-m}^{r,c,j'} A_{c-p'}^{r,c,j'} e^{-i\lambda_{r,c,j}t} e^{i\lambda_{r,c,j'}t} \langle c-p|\rho_f(0)|c-p'\rangle \langle r,p|\rho_{TLM}(0)|r,p'\rangle,
\end{aligned} \tag{18}$$

As is seen, these equations are quite complicated and to use the general expressions would be prohibitive except for the simplest of cases. Equations (18) could possibly be rewritten in terms of sin's and cos's by taking advantage of the symmetry about "c" in the sums over j and j'. It is noted that some simplification may be possible by retaining only terms which have the "slowest" time dependence. This is possible because of the orthogonality between states of different j and/or n subscripts. This completes the formulation of the expressions. We consider one special case next.

---

[8] The factors $\left(\frac{\gamma}{\mu}\right)$ and $\left|\frac{\gamma}{\mu}\right|^2$ seen in eq. (18) differ from those same constants if Reference 3 by a factor of 2, i.e. $\gamma$ is $2\gamma$ in that reference. We ignore the difference here.

## All TLMs in the up State at time=0

For this special case $r = \frac{N}{2}$ and $P(r) = 1$. Further

$$\langle r,p|\rho_{TLM}(0)|r,p'\rangle = \delta_{p,\frac{N}{2}}\delta_{p',\frac{N}{2}} \qquad (19)$$

Equation (18) becomes

$$\langle E^-(t)\rangle = -\left(\frac{\gamma}{\mu}\right)\sum_{m=-\frac{N}{2}}^{\frac{N}{2}}\sum_{c=m}^{\infty}(c-m+1)^{1/2}\langle c-\frac{N}{2}|\rho_f(0)|c+1-\frac{N}{2}\rangle$$

$$\times \sum_{j=0}^{Min[N,c+\frac{N}{2}]}\sum_{j'=0}^{Min[N,c+1+\frac{N}{2}]} e^{-i\left(\lambda_{\frac{N}{2},c,j}-\lambda_{\frac{N}{2},c+1,j'}\right)t}(A^*)^{\frac{N}{2},c+1,j}_{c+1-\frac{N}{2}}A^{\frac{N}{2},c,j}_{c-m}(A^*)^{\frac{N}{2},c,j'}_{c-m}A^{\frac{N}{2},c+1,j'}_{c+1-\frac{N}{2}}$$

$$\langle E^-E^+(t)\rangle = \left|\frac{\gamma}{\mu}\right|^2 \sum_{m=-\frac{N}{2}}^{\frac{N}{2}}\sum_{c=m}^{\infty}(c-m)\langle c-\frac{N}{2}|\rho_f(0)|c-\frac{N}{2}\rangle$$

$$\times \sum_{j=0}^{Min[N,c+\frac{N}{2}]}\sum_{j'=0}^{Min[N,c+\frac{N}{2}]} e^{-i\left(\lambda_{\frac{N}{2},c,j}-\lambda_{\frac{N}{2},c,j'}\right)t}(A^*)^{\frac{N}{2},c,j}_{c-\frac{N}{2}}A^{\frac{N}{2},c,j}_{c-m}(A^*)^{\frac{N}{2},c,j'}_{c-m}A^{\frac{N}{2},c,j'}_{c-\frac{N}{2}}$$

(20)

In addition, the density matrices $\langle c-\frac{N}{2}|\rho_f(0)|c-\frac{N}{2}\rangle$ and $\langle c-\frac{N}{2}|\rho_f(0)|c+1-\frac{N}{2}\rangle$ only have values for $c-\frac{N}{2} \geq 0$. This allows for a further reduction of eq. (20)

$$\langle E^-(t)\rangle = -\left(\frac{\gamma}{\mu}\right)\sum_{m=-\frac{N}{2}}^{\frac{N}{2}}\sum_{c=\frac{N}{2}}^{\infty}(c-m+1)^{1/2}\langle c-\frac{N}{2}|\rho_f(0)|c+1-\frac{N}{2}\rangle$$

$$\times \sum_{j=0}^{N}\sum_{j'=0}^{N} e^{-i\left(\lambda_{\frac{N}{2},c,j}-\lambda_{\frac{N}{2},c+1,j'}\right)t}(A^*)^{\frac{N}{2},c+1,j}_{c+1-\frac{N}{2}}A^{\frac{N}{2},c,j}_{c-m}(A^*)^{\frac{N}{2},c,j'}_{c-m}A^{\frac{N}{2},c+1,j'}_{c+1-\frac{N}{2}}$$

(21)

$$\langle E^-E^+(t)\rangle = \left|\frac{\gamma}{\mu}\right|^2 \sum_{m=-\frac{N}{2}}^{\frac{N}{2}}\sum_{c=\frac{N}{2}}^{\infty}(c-m)\langle c-\frac{N}{2}|\rho_f(0)|c-\frac{N}{2}\rangle$$

$$\times \sum_{j=0}^{N}\sum_{j'=0}^{N} e^{-i\left(\lambda_{\frac{N}{2},c,j}-\lambda_{\frac{N}{2},c,j'}\right)t}(A^*)^{\frac{N}{2},c,j}_{c-\frac{N}{2}}A^{\frac{N}{2},c,j}_{c-m}(A^*)^{\frac{N}{2},c,j'}_{c-m}A^{\frac{N}{2},c,j'}_{c-\frac{N}{2}}$$

The first order correlation function as a function of time can be simplified by using the summation variables m=p - $\frac{N}{2}$ and c=n+$\frac{N}{2}$ and using the orthogonality relationships for the $A_n^{r,c,j}$ and thus obtaining

$$\langle E^- E^+(t) \rangle = \left|\frac{\gamma}{\mu}\right|^2 [\bar{n} + S_1(\bar{n}, N, \gamma t)], \qquad (22)$$

where $\bar{n}$ is the mean number of photons and

$$S_1(\bar{n}, N, \gamma t) = \sum_{n=0}^{\infty} \langle n|\rho_f(0)|n\rangle \sum_{j=0}^{N} \sum_{j'=0}^{N} e^{-i\left(\lambda_{\frac{N}{2},(n+\frac{N}{2}),j} - \lambda_{\frac{N}{2},(n+\frac{N}{2}),j'}\right)t} (A^*)_n^{\frac{N}{2},(n+\frac{N}{2}),j} A_n^{\frac{N}{2},(n+\frac{N}{2}),j'}$$
$$\times \sum_{p=0}^{N} p\, A_{n+p}^{\frac{N}{2},(n+\frac{N}{2}),j} (A^*)_{n+p}^{\frac{N}{2},(n+\frac{N}{2}),j'} \qquad (23)$$

By scuffling terms and using orthogonality relationships again

$$S_1(\bar{n}, N, \gamma t) = -4 \sum_{n=0}^{\infty} \langle n|\rho_f(0)|n\rangle \sum_{j=0}^{N-1} \sum_{j'=j+1}^{N} \mathrm{Sin}^2\left\{\frac{\left[q_{\frac{N}{2},(n+\frac{N}{2}),j} - q_{\frac{N}{2},(n+\frac{N}{2}),j'}\right]}{2}\Omega|\kappa|t\right\}$$
$$\times (A^*)_n^{\frac{N}{2},(n+\frac{N}{2}),j} A_n^{\frac{N}{2},(n+\frac{N}{2}),j'} \sum_{p=0}^{N} p\, A_{n+p}^{\frac{N}{2},(n+\frac{N}{2}),j} (A^*)_{n+p}^{\frac{N}{2},(n+\frac{N}{2}),j'} \qquad (24)$$

Equation (24) is the final general solution for $S_1(\bar{n}, N, \gamma t)$ and we have used the alternate expression for the effective eigenvalues of reference (6), namely $\lambda = c - |\kappa| q$ and multiplied by $\Omega$ to obtain the correct units even though $\hbar$ is still set to unity. Note that $\Omega|\kappa|t = \gamma t$. Using the notation introduced by Cummings (1), the ensemble average for the field $\langle E^-(t) \rangle$ can be written as

$$\langle E^-(t) \rangle = -\left(\frac{\gamma}{\mu}\right) e^{i\omega t} S_2(\bar{n}, N, \gamma t) \qquad (25)$$

where

$$S_2(\bar{n}, N, \gamma t) = \sum_{n=0}^{\infty} \langle n|\rho_f(0)|n+1\rangle \sum_{k=0}^{N} (n+k+1)^{1/2}$$
$$\times \left\{ \sum_{j=0}^{N} \sum_{j'=0}^{N} \mathrm{Cos}\left[\left(q_{\frac{N}{2},n+\frac{N}{2},j} - q_{\frac{N}{2},n+\frac{N}{2}+1,j'} - \beta\right)\Omega|\kappa|t\right] (A^*)_{n+1}^{\frac{N}{2},n+\frac{N}{2}+1,j} A_{n+k}^{\frac{N}{2},n+\frac{N}{2},j} (A^*)_{n+k}^{\frac{N}{2},n+\frac{N}{2},j'} A_{n+1}^{\frac{N}{2},n+\frac{N}{2}+1,j'} \right.$$
$$\left. + i \sum_{j=0}^{N} \sum_{j'=0}^{N} \mathrm{Sin}\left[\left(q_{\frac{N}{2},n+\frac{N}{2},j} - q_{\frac{N}{2},n+\frac{N}{2}+1,j'} - \beta\right)\Omega|\kappa|t\right] (A^*)_{n+1}^{\frac{N}{2},n+\frac{N}{2}+1,j} A_{n+k}^{\frac{N}{2},n+\frac{N}{2},j} (A^*)_{n+k}^{\frac{N}{2},n+\frac{N}{2},j'} A_{n+1}^{\frac{N}{2},n+\frac{N}{2}+1,j'} \right\} \qquad (26)$$

and we have taken advantage of the real nature of the $A_n^j$. I have not been able to reduce eq.24 and 26 to simpler expressions except for resonance ($\beta = 0$).

## Resonance

For resonance eq. 26 can be further reduced. After considerable algebra and making use of symmetry for the exact eigenvectors and effective eigenvalues, it can be shown that the imaginary part of eq. 26 is zero. This means that

$$S_2(\bar{n}, N, \gamma t) = \sum_{n=0}^{\infty} \langle n | \rho_f(0) | n+1 \rangle \sum_{k=0}^{N} (n+k+1)^{1/2}$$
$$\times \sum_{j=0}^{N} \sum_{j'=0}^{N} \left\{ Cos\left[q_{\frac{N}{2},n+\frac{N}{2},j} \Omega |\kappa| t\right] Cos\left[q_{\frac{N}{2},n+\frac{N}{2}+1,j'} \Omega |\kappa| t\right] + Sin\left[q_{\frac{N}{2},n+\frac{N}{2},j} \Omega |\kappa| t\right] Sin\left[q_{\frac{N}{2},n+\frac{N}{2}+1,j'} \Omega |\kappa| t\right] \right\} \quad (27)$$
$$\times (A^*)_{n+1}^{\frac{N}{2},n+\frac{N}{2}+1,j} A_{n+k}^{\frac{N}{2},n+\frac{N}{2},j} (A^*)_{n+k}^{\frac{N}{2},n+\frac{N}{2},j'} A_{n+1}^{\frac{N}{2},n+\frac{N}{2}+1,j'} \quad \text{for } \beta = 0$$

At this stage in the simplification, we see some similarity to the Cumming's expression for $S_2$. There is some further reduction possible if one focuses on odd or even numbers of TLMs. We only do that for eq.27. Thus for $\beta = 0$ and using $A_{n+k}^{\frac{N}{2},n+\frac{N}{2},N-j} = (-1)^k A_{n+k}^{\frac{N}{2},n+\frac{N}{2},j}$

$$S_2(\bar{n}, N, \gamma t) = \frac{1}{2} \sum_{n=0}^{\infty} \langle n | \rho_f(0) | n+1 \rangle \sum_{k=0}^{N} (n+k+1)^{1/2}$$
$$\times \left[ [1+(-1)^k] \left\{ \sum_{j=0}^{N} Cos\left[q_{\frac{N}{2},n+\frac{N}{2},j} \Omega |\kappa| t\right] (A^*)_{n+1}^{\frac{N}{2},n+\frac{N}{2}+1,j} A_{n+k}^{\frac{N}{2},n+\frac{N}{2},j} \right\} \left\{ \sum_{j'=0}^{N} Cos\left[q_{\frac{N}{2},n+\frac{N}{2}+1,j'} \Omega |\kappa| t\right] (A^*)_{n+1}^{\frac{N}{2},n+\frac{N}{2}+1,j'} A_{n+k}^{\frac{N}{2},n+\frac{N}{2},j'} \right\} \right. \quad (28)$$
$$\left. + [1-(-1)^k] \left\{ \sum_{j=0}^{N} Sin\left[q_{\frac{N}{2},n+\frac{N}{2},j} \Omega |\kappa| t\right] (A^*)_{n+1}^{\frac{N}{2},n+\frac{N}{2}+1,j} A_{n+k}^{\frac{N}{2},n+\frac{N}{2},j} \right\} \left\{ \sum_{j'=0}^{N} Sin\left[q_{\frac{N}{2},n+\frac{N}{2}+1,j'} \Omega |\kappa| t\right] (A^*)_{n+1}^{\frac{N}{2},n+\frac{N}{2}+1,j'} A_{n+k}^{\frac{N}{2},n+\frac{N}{2},j'} \right\} \right]$$

This implies that only alternating terms in k exist for the Cosines and Sines.

We now move to examples for various values of N.

### 1 TLM

For N=1 and resonance, the expressions for $S_2(\bar{n}, 1, \gamma t)$ and $S_1(\bar{n}, 1, \gamma t)$ are given by

$$S_2(\bar{n}, 1, \gamma t) = \sum_{n=0}^{\infty} \langle n | \rho_f(0) | n+1 \rangle \{ \sqrt{n+1} Cos[\sqrt{n+1} \Omega |\kappa| t] Cos[\sqrt{n+2} \Omega |\kappa| t] \quad (29a)$$
$$+ \sqrt{n+2} Sin[\sqrt{n+1} \Omega |\kappa| t] Sin[\sqrt{n+2} \Omega |\kappa| t] \} : N = 1, \beta = 0$$

$$S_1(\bar{n}, 1, \gamma t) = \sum_{n=0}^{\infty} \langle n | \rho_f(0) | n \rangle Sin^2[\sqrt{n+1} \Omega |\kappa| t] : N = 1, \beta = 0 \quad (29b)$$

These results agree with those found by Cummings.

## 2 TLMs

For N=2 and $\beta = 0$, $S_1(\bar{n},2,\gamma t)$ and $S_2(\bar{n},2,\gamma t)$ are given by;

$$S_1(\bar{n},2,\gamma t) = 8\sum_{n=0}^{\infty}\langle n|\rho_f(0)|n\rangle\left\{Sin^2\left[\sqrt{n+\frac{3}{2}}\Omega|\kappa|t\right]\frac{(n+1)(n+2)}{(2n+3)^2}\right.$$
$$\left.-\frac{1}{8}Sin^2\left[2\sqrt{n+\frac{3}{2}}\Omega|\kappa|t\right]\frac{(n+1)}{(2n+3)^2}\right\} \quad (30a)$$

$$S_2(\bar{n},2,\gamma t) = \sum_{n=0}^{\infty}\frac{\langle n|\rho_f(0)|n+1\rangle(n+2)}{(2n+3)(2n+5)}$$
$$\times\left\{[(n+1)\sqrt{n+1}+(n+2)\sqrt{n+3}]Cos(\sqrt{2}\sqrt{2n+3}\Omega|\kappa|t)Cos(\sqrt{2}\sqrt{2n+5}\Omega|\kappa|t)\right.$$
$$+[(n+1)\sqrt{3+n}-(n+3)\sqrt{n+1}][Cos(\sqrt{2}\sqrt{2n+3}\Omega|\kappa|t)+Cos(\sqrt{2}\sqrt{2n+5}\Omega|\kappa|t)]$$
$$\left.+(n+3)\left[\sqrt{n+1}+\frac{(n+1)\sqrt{n+3}}{(n+2)}\right]+(2n+3)\sqrt{n+2}Sin[\sqrt{2}\sqrt{2n+3}\Omega|\kappa|t]Sin[\sqrt{2}\sqrt{2n+5}\Omega|\kappa|t]\right\} \quad (30b)$$

We note the existence of a constant term for an even number of TLMs. In Appendix C, we present the analytic results for 3 TLMs and 4 TLMs for the resonant case.

## Non Resonance

### 1 TLM Non Resonant

For N=1 and non resonance, the expressions for $S_1(\bar{n},1,\gamma t)_{NR}$ and $S_2(\bar{n},1,\gamma t)_{NR}$ are given by

$$S_1(\bar{n},1,\gamma t)_{NR} = \sum_{n=0}^{\infty}\langle n|\rho_f(0)|n\rangle\left(\frac{n+1}{n+1+\Delta}\right)Sin^2[\sqrt{n+1+\Delta}\Omega|\kappa|t]$$
$$\Delta = \frac{\beta^2}{4}, N = 1 \quad (31a)$$

$$S_2(\bar{n},1,\gamma t)_{NR} = \sum_{n=0}^{\infty} \frac{\langle n|\rho_f(0)|n+1\rangle}{4\sqrt{n+1+\Delta}\sqrt{n+2+\Delta}}$$
$$\times \left\{\left[-\left(\left[(-\sqrt{1+n}+\sqrt{2+n})\sqrt{\Delta}+\sqrt{(1+n)(n+1+\Delta)}\right.\right.\right.\right.$$
$$+\sqrt{(2+n)(n+1+\Delta)}\right]$$
$$\times(\sqrt{\Delta}-\sqrt{n+2+\Delta})\cos[(\sqrt{n+1+\Delta}-\sqrt{n+2+\Delta})\Omega\kappa t]\right)$$
$$+\left((\sqrt{1+n}-\sqrt{2+n})\sqrt{\Delta}+\sqrt{(1+n)(n+1+\Delta)}+\sqrt{(2+n)(n+1+\Delta)}\right) \quad (31b)$$
$$\times(\sqrt{\Delta}+\sqrt{n+2+\Delta})\cos[(\sqrt{n+1+\Delta}-\sqrt{n+2+\Delta})\Omega\kappa t]$$
$$+8\sqrt{n+2}(1+n-\sqrt{2+3n+n^2})\cos[(\sqrt{n+1+\Delta}+\sqrt{n+2+\Delta})\Omega\kappa t]\right]$$
$$+2i\sqrt{\Delta}\left[(\sqrt{(1+n)(n+1+\Delta)}+\sqrt{(2+n)(n+1+\Delta)}\right.$$
$$+\sqrt{(1+n)(n+2+\Delta)}$$
$$-\sqrt{(2+n)(n+2+\Delta)})\sin[(\sqrt{n+1+\Delta}-\sqrt{n+2+\Delta})\Omega\kappa t]\right]\right\}: N=1$$

We see here, confirmation that for non resonance $S_2$ is in fact complex. We do not present analytic results for non resonance for greater than 1 TLM. Instead in Appendix D we briefly discuss the reason for this.

# Discussion

In this paper, we have continued the exploration of the formulation for the ensemble averages of $E^-$ and $E^-E^+$ for N-TLMs first discussed in Appendix F of the second citation for reference 2. In the original work, the general solutions were provided and derivations for all TLMs in the up state were also expressed. Here, the work is carried out much further using orthogonality relationships and Sin and Cosine functional expansions, with the final results being the expressions for $S_1$ used in $E^-E^+$ and $S_2$ used in $E^-$.

To continue with the analysis, approximations can be made for values of n$\gg N$. Since it will be difficult to use the general expressions (Eqs. 24 and 25), we turn to the specific examples for the resonant cases for N$>$ 1. For N=2, the $S_1(\bar{n}, 2, \gamma t) \to 2S_1(\bar{n}, 1, \gamma t)$ with the second term in $S_1(\bar{n}, 2, \gamma t)$ decaying as $\frac{1}{n}$ as n becomes larger. In a similar fashion for N=3, $S_1(\bar{n}, 3, \gamma t) \to 3S_1(\bar{n}, 1, \gamma t)$ again with the other terms decaying like $\frac{1}{n}$. This leads one to speculate that, for a large number of photons, the individual TLMs act independently. Recall that in Reference 5, one of us, developed an approximation (for the resonant case) valid when c>>r, the so called Average Field Approximation (AFA). In Appendix A, we develop the same approximation but for the non-resonant case. Secondly, in Appendix B, we apply the AFA to the evaluation of $S_1(\bar{n}, N, \gamma t)$ and $S_2(\bar{n}, N, \gamma t)$ and perform some numerical justification for the use of this approximation. As can be seen by reference to Eqs. B8 and B15, the analytic expressions are quite simple and justify the approximation.

## Numerical Evaluation of $S_1(\bar{n}, N, \gamma t)$.

The exact calculation for this expression is performed using eq. 24 while the AFA approximation is obtained from eq. B.8. Several examples are presented and it is found that if the average photon number is reasonably large and the photon number density is small near n=0, than the two results for resonance are nearly indistinguishable. On the other hand for non-resonance, even though the overall mean and position of oscillation are the same, the details of each oscillation spike can be considerably different. In order to demonstrate this, we present 4 figures for the coherent photon number density function

$$\langle n|\rho_f(0)|n\rangle = \frac{e^{-\bar{n}}\bar{n}^n}{n!} \tag{32}$$

for $\bar{n} = 100$ and the number of TLMs=4.

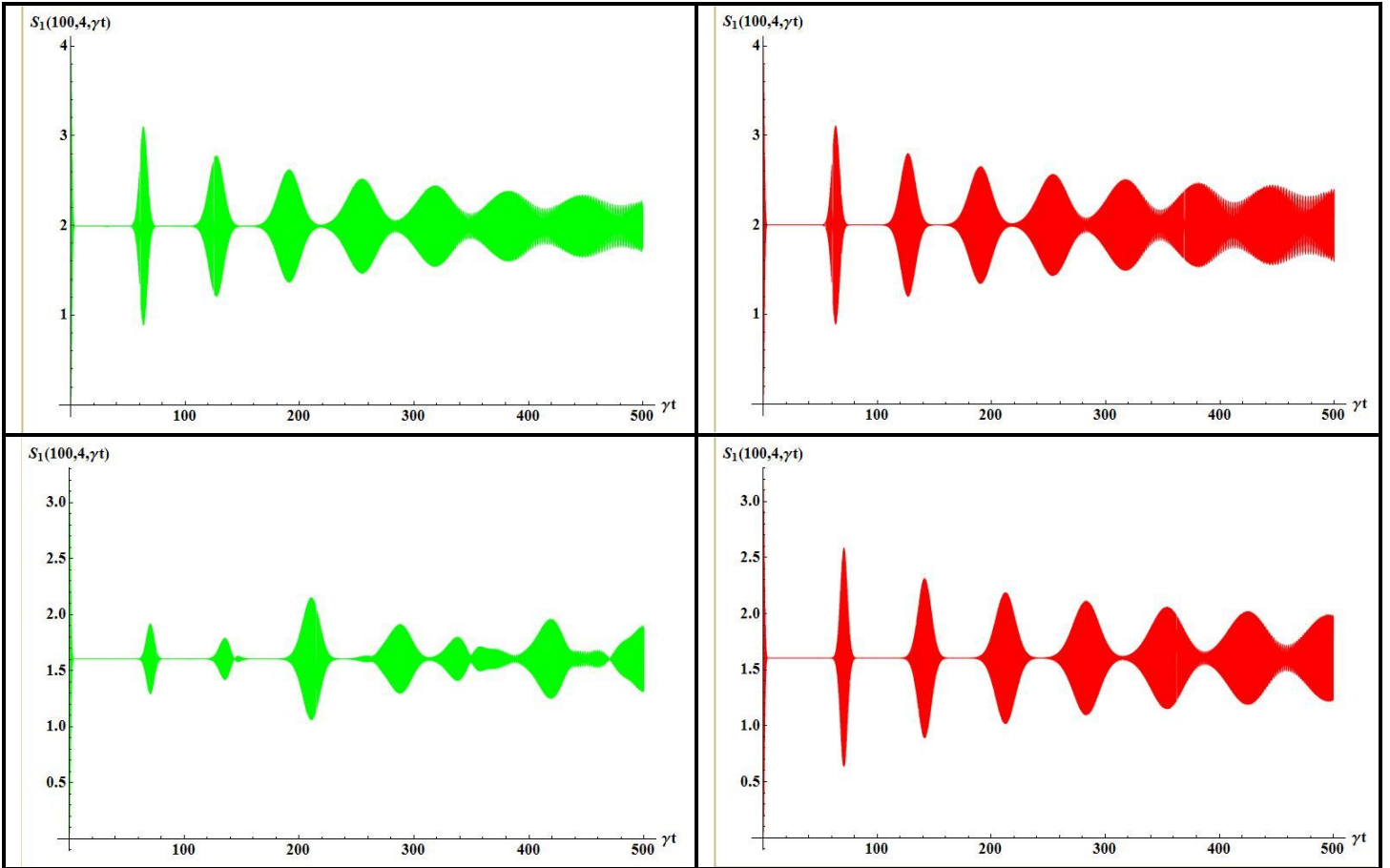

**Figure 1: Numerical Solutions for $S_1(\bar{n}, N, \gamma t)$ for 4 TLMs and $\bar{n} = 100$. The upper curves are for the resonant case while the lower curves have a non resonance parameter of $\Delta = 25$. The curves on the left are the exact solutions while those on the right are for the Average Field Approximation.**

If one were to examine the individual oscillations, one would see that the oscillations are shifted slightly between the exact and AFA. An interesting observation is that as the mean number on photons decreases, so does the mean of $S_1$. For instance for $\bar{n} = 1$, the mean of $S_1$ is 1.66. For $\bar{n} = 4$ the mean for $S_1$ is 1.84. Even for $\bar{n} = 10$ the mean for $S_1$ is 1.93. This means that unless the average number of photons is reasonably larger than the number of TLMs, the stimulated emission is not sufficient to dump half the TLM photon energy.

For a thermal distribution

$$\langle n|\rho_f(0)|n\rangle = \frac{1}{1+\bar{n}}\left(\frac{\bar{n}}{1+\bar{n}}\right)^n \tag{33}$$

the number of photons peaks near n=0. For this reason the exact and AFA do not agree well for resonance although again the mean value for $S_1$ agree for the exact and AFA. For non resonance, an example has been provided.

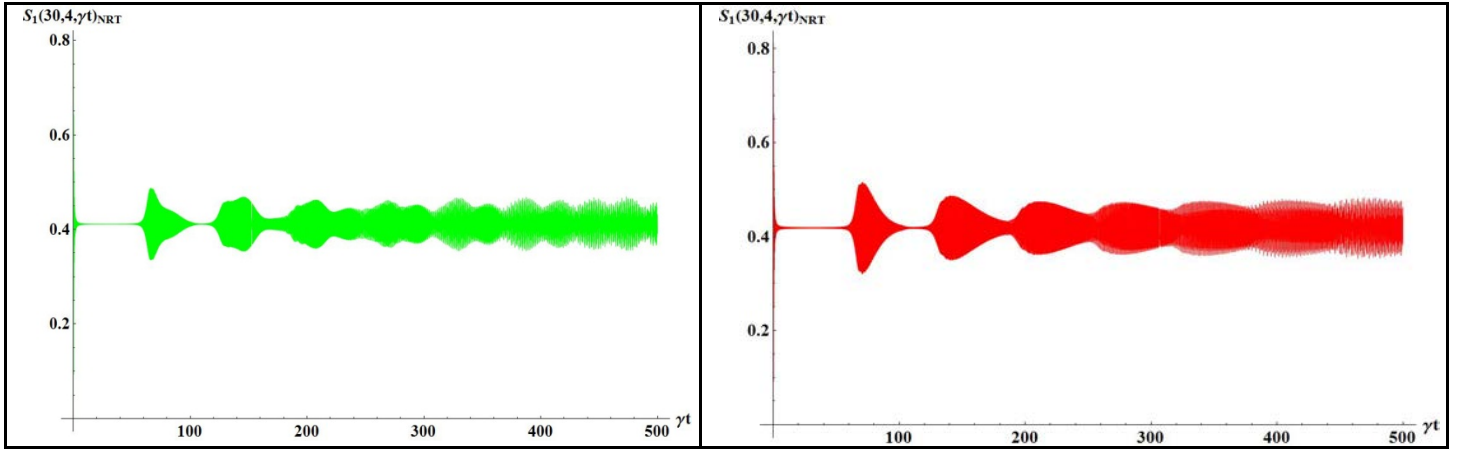

**Figure 2: Numerical Solutions for $S_1(\bar{n}, N, \gamma t)$ for 4 TLMs and $\bar{n} = 30$. These 2 curves have a non resonance parameter of $\Delta = 100$. The curves on the left are the exact solutions while those on the right are for the Average Field Approximation. Both are calculated assuming a thermal distribution for the photon number density.**

As for the case for the coherent number distribution, the exact and AFA results are alike in form but different in detail.

For further examples for which not all TLMs are in the up state, see Appendices E and F. In appendix E, cases where all TLMs are down except for 1 or 2 TLMs which are up and there are no initial photons in the field are examined. Spontaneous Emission suppression is shown. In addition, if half the TLMs are in the up state, Super Radiant behavior is seen. Finally we will see in Appendix F, stimulated absorption when all the TLMs are initially in the down state.

As has been seen, the use of the Eigenvectors for the Hamiltonian in eq. (6) developed in Reference (6) has proven useful as an expansion basis for this class of problem.

## Appendix A: AFA for Resonance and Non Resonance

As for the resonance case discussed in the thesis (Ref. 2), the eigenvectors $|r, c, j>_{AFA}$ may be expressed in terms of the upper and lower states of a single TLM as

$$|r,c,j>_{AFA} = \frac{1}{\sqrt{\frac{N!}{n_+! n_-!}}} \sum_\rho \left\{ \prod_{k=1}^{n_+} \left( b1|c_j - \frac{1}{2}> |\frac{1}{2}, \frac{1}{2}> + b2|c_j + \frac{1}{2}> |\frac{1}{2}, -\frac{1}{2}> e^{i(\pi - \varphi)} \right)_k \right.$$
$$\left. \times \prod_{k'=1}^{n_-} \left( a1|c_j - \frac{1}{2}> |\frac{1}{2}, \frac{1}{2}> e^{i\varphi} + a2|c_j + \frac{1}{2}> |\frac{1}{2}, -\frac{1}{2}> \right)_{k'} \right\}, r = \frac{N}{2}$$

(A.1)

where $\rho$ represents the various permutations on the ground and excited states and k and $k'$ represent the different excited and ground state TLMs interacting with the field and (a1,a2,b1,b2) are the coefficients of the TLM upper and lower states defined in terms of the molecular up and down states (the sign of b2 has been accounted for in the phase factor of A.1). This state may be represented by a double sum

$$|r,c,j>_{AFA}= \frac{1}{\sqrt{\frac{N!}{n_+!n_-!}}}\sum_{k=0}^{n_+}\sum_{k'=0}^{n_-}(b1)^{n_+-k}(b2)^k(a1)^{n_--k'}(a2)^{k'}e^{ik(\pi-\varphi)}e^{i(n_--k')\varphi}$$

$$\times \left[\frac{n_+!}{k!\,(n_+-k)!}\right] \times \left[\frac{n_-!}{k'!\,(n_--k')!}\right] \times \left[\frac{N!}{n_+!n_-!}\right] \div \left[\frac{N!}{(k+k')!\,(N-k-k')!}\right] \quad (A.2)$$

$$\times \sqrt{\frac{N!}{(k+k')!\,(N-k-k')!}}\,|c-(r-(k+k'))>|r,r-(k+k')>$$

Following the same reasoning as (Ref.2)

$$|\frac{N}{2},c,j>_{AFA}= \frac{(b1)^j(a1)^{N-j}}{\sqrt{\frac{N!}{n_+!n_-!}}}\sum_{L=0}^{N}\sum_{L'=down}^{up}{}'\left(\frac{b2a2}{b1a1}\right)^{\frac{L}{2}}\left(\frac{b2a1}{b1a2}\right)^{\frac{L'}{2}}e^{\frac{(L+L')\pi i}{2}}[L!\,(2r-L)!]$$

$$\div \left[\left(\frac{L+L'}{2}\right)!\left(\frac{L-L'}{2}\right)!\left(j-\left(\frac{L+L'}{2}\right)\right)!\left(n_--\left(\frac{L-L'}{2}\right)\right)!\right]$$

$$\times e^{(N-j-L)\varphi}\sqrt{\frac{(2r)!}{L!\,(2r-L)!}}\,|c-m>|r,m>,$$

where
$$(A.3)$$

$$n_+ = j$$
$$n_- = N-j,$$
$$m = \frac{N}{2}-L,$$
$$down = max[-L, L-2n_-],$$
$$up = min[L, 2j-L].$$

The prime on the second sum indicates that only every other value starting with "down" is used in the sum. We already know that the eigenvectors can be expressed in the form ($r = \frac{N}{2}$ and $c$ is choosen as $n+\frac{N}{2}$)

$$|\frac{N}{2},c,j> = \sum_{p=0}^{N}A_p^{\left(\frac{N}{2},n+\frac{N}{2},j\right)}|n+p>|\frac{N}{2},\frac{N}{2}-p>, \text{ when } n>r \quad (A.4)$$

Making the correspondence to eq. A.3, we can perform the substitution p=L and

$$down = \begin{matrix} -p & j < N - p \\ p + 2j - 2N & j \geq N - p \end{matrix}$$
$$up = \begin{matrix} p & j \geq p \\ 2j - p & j < p \end{matrix}$$
(A.5)

express $A_{n+p}^{(r,c,j)}$ after some algebraic manipulation as

$$A_{n+p}^{(\frac{N}{2},n+\frac{N}{2},j)} = (b1)^j (a1)^{N-j} \begin{cases} \left(\frac{a2}{a1}\right)^p \sqrt{\frac{(N-j)!(N-p)!}{p!j!}} \frac{F\left(-j,-p,N+1-p-j:-\left(\frac{b2a1}{b1a2}\right)\right)}{(N-j-p)!} & N \geq j+p \\ (-1)^{p-N} \left(\frac{b2a2}{b1a1}\right)^p \left(\frac{b2a1}{b1a2}\right)^{-N} \sqrt{\frac{p!j!}{(N-p)!(N-j)!}} \frac{F\left(j-N,p-N,1+p+j-N:-\left(\frac{b2a1}{b1a2}\right)\right)}{(j+p-N)!} & N < j+p \end{cases}$$
(A.6)

where F is the Hypergeometric function. The quantities. (a1,a2,b1,b2) are the coefficients for the exact solution for the single TLM in the AFA and are given by

$$a1 = \frac{1}{\sqrt{1 + [0.5\beta - 0.5\sqrt{4. + \beta^2}]^2}} \quad a2 = \frac{-0.5\beta + 0.5\sqrt{4. + \beta^2}}{\sqrt{1 + [0.5\beta - 0.5\sqrt{4. + \beta^2}]^2}}$$
$$b1 = \frac{1}{\sqrt{1 + [0.5\beta + 0.5\sqrt{4. + \beta^2}]^2}} \quad b2 = \frac{0.5\beta + 0.5\sqrt{4. + \beta^2}}{\sqrt{1 + [0.5\beta + 0.5\sqrt{4. + \beta^2}]^2}}$$
(A.7)

Further the effective eigenvalues in the AFA lie along a straight line and are given by

$$\bar{q}_j = -\left(\frac{N}{2} + n\right)\bar{\beta} + (r - j)\sqrt{4 + \bar{\beta}^2}$$
(A.8)

and

$$\bar{\beta} = \frac{\omega - \Omega}{|\kappa|\Omega\sqrt{n_o}}$$
(A.9)

# Appendix B: The AFA applied to the solution of $S_2(\bar{n}, N, \gamma t)$ and $s_2(\bar{n}, N, \gamma t)$

In this appendix, we apply the AFA to simplify Eqs. 24 and 26.

## Evaluation of $S_1(\bar{n}, N, \gamma t)$ using the Average Field Approximation (AFA)

Apply the AFA to eq. 24 with the following notation

$$S_1(\bar{n}, N, \gamma t)_{AFA} = -4 \sum_{n=0}^{\infty} \langle n | \rho_f(0) | n \rangle$$

$$\times \sum_{j=0}^{N-1} \sum_{j'=j+1}^{N} Sin^2 \left\{ \frac{\left[ q_{\frac{N}{2},(n+\frac{N}{2}),j} - q_{\frac{N}{2},(n+\frac{N}{2}),j'} \right]}{2} \Omega |\kappa| t \right\} (A^*)_n^{\frac{N}{2},(n+\frac{N}{2}),j} A_n^{\frac{N}{2},(n+\frac{N}{2}),j'} \quad \text{(B.1)}$$

$$\times \sum_{p=0}^{N} p A_{n+p}^{\frac{N}{2},(n+\frac{N}{2}),j} (A^*)_{n+p}^{\frac{N}{2},(n+\frac{N}{2}),j'}$$

The Sin function reduces to $Sin^2 \left[ (j' - j) \sqrt{n_o} \sqrt{1 + \frac{\beta^2}{4}} \Omega |\kappa| t \right]$ due to eq. (A.8). This means that the expression can be rearranged for the differences in j'-j from 1 to N. We know from eq. (A.6) that

$$A_n^{\frac{N}{2},(n+\frac{N}{2}),j} = (b1)^j (a1)^{N-j} \sqrt{\frac{N!}{j!(N-j)!}} \quad \text{(B.2)}$$

The evaluation of $\sum_{p=0}^{N} p A_{n+p}^{\frac{N}{2},(n+\frac{N}{2}),j} (A^*)_{n+p}^{\frac{N}{2},(n+\frac{N}{2}),j'}$ appears to be exceedingly complex. Fortunately, we do not have to use eq. (A.6) for that evaluation. Instead we note that the Hamiltonian for the AFA is tri-diagonal with $c - \frac{\omega - \Omega}{\Omega} n$ along the diagonal and the off diagonal terms $= -\sqrt{n_0}\sqrt{r(r+1) - m(m+1)} |\kappa|$ with $r \geq m \geq -r$. In finding the eigenvalues for this Hamiltonian, the unknown eigenvalue $\lambda$ is subtracted from the diagonal elements. Then we divide this resulting Hamiltonian by $\sqrt{n_0} |\kappa|$. The effective eigenvalues of q lie along the resulting diagonal. This is the actual matrix used to determine the eigenvalues and vectors. The transformation used to diagonalize the resulting matrix is

$$\bar{A} \, \bar{H} \bar{A}^\dagger = \bar{E}, \quad \text{(B.3)}$$

where $\bar{E}$ is the eigenvalue matrix with the eigenvalues along the diagonal and $\bar{A}$ is the matrix of the eigenvectors. Inverting this expression to solve for H yields

$$\bar{H} = \bar{A}^\dagger \bar{E} \bar{A} = \bar{A}^\dagger \bar{\lambda} \bar{A}$$

$$= \left\{ -N\sqrt{1+\frac{\bar{\beta}^2}{4}}\sqrt{n_o}|\kappa| + \left(\frac{N}{2}+n\right)(1+\beta|\kappa|) \right\} \bar{I} + 2\sqrt{1+\frac{\bar{\beta}^2}{4}}\sqrt{n_o}|\kappa| \sum_{p=0}^{N} p\, A_{n+p}^{\frac{N}{2},\left(n+\frac{N}{2}\right),j} A_{n+p}^{\frac{N}{2},\left(n+\frac{N}{2}\right),j'} \quad (B.4)$$

where $\bar{I}$ is the identity matrix.

If we now equate terms, we see that $\sum_{p=0}^{N} p\, A_{n+p}^{\frac{N}{2},\left(n+\frac{N}{2}\right),j} (A^*)_{n+p}^{\frac{N}{2},\left(n+\frac{N}{2}\right),j'}$ is only non zero along the diagonal or when j'=±j. In fact we find that

$$\sum_{p=0}^{N} p\, A_{n+p}^{\frac{N}{2},\left(n+\frac{N}{2}\right),j} (A^*)_{n+p}^{\frac{N}{2},\left(n+\frac{N}{2}\right),j+1} = -\frac{\sqrt{(j+1)(N-j)}}{2\sqrt{1+\frac{\beta^2}{4}}} \delta_{j,j+1} \quad (B.5)$$

From eq. (B.2)

$$(A^*)_n^{\frac{N}{2},\left(n+\frac{N}{2}\right),j} A_n^{\frac{N}{2},\left(n+\frac{N}{2}\right),j'} \sum_{p=0}^{N} p\, A_{n+p}^{\frac{N}{2},\left(n+\frac{N}{2}\right),j} (A^*)_{n+p}^{\frac{N}{2},\left(n+\frac{N}{2}\right),j'} = N(a1b1)\frac{(N-1)![(b1)^2]^j[(a1)^2]^{N-j-1}}{2j!(N-1-j)!\sqrt{1+\frac{\beta^2}{4}}}\delta_{j,j+1} \quad (B.6)$$

Performing the summation over j from 0 to N-1 and substituting the values for (a1,b1) yields $\frac{N}{4\left(1+\frac{\beta^2}{4}\right)}$. Thus

$$S_1(\bar{n}, N, \gamma t)_{AFA} = N \sum_{n=0}^{\infty} \langle n|\rho_f(0)|n\rangle \frac{n_0}{n_0+\Delta} Sin^2\left[\sqrt{n_0 + \Delta\Omega}|\kappa|t\right] \quad (B.7)$$

Note that this expression has the complete expression for $\beta$ and $\Delta = \frac{\beta^2}{4}$. If we replace $n_o$ by c=n+$\frac{N}{2}$, the final results becomes

$$S_1(\bar{n}, N, \gamma t)_{AFA} = N \sum_{n=0}^{\infty} \langle n|\rho_f(0)|n\rangle \frac{n+\frac{N}{2}}{n+\frac{N}{2}+\Delta} Sin^2\left[\sqrt{n+\frac{N}{2}+\Delta\Omega}|\kappa|t\right] \quad (B.8)$$

This is the non resonant results. The resonant results is obtained by setting $\Delta = 0$.

## Evaluation of $S_2(\bar{n}, N, \gamma t)$ using the Average Field Approximation (AFA)

Unfortunately the evaluation of $S_2(\bar{n}, N, \gamma t)_{AFA}$ is not quite as straightforward and we need to make a further assumption. In this case the quantity from eq. 26 as follows

$$(A^*)_{n+1}^{\frac{N}{2},n+\frac{N}{2}+1,j} A_{n+1}^{\frac{N}{2},n+\frac{N}{2}+1,j'} \sum_{p=0}^{N} (n+p+1)^{1/2} A_{n+p}^{\frac{N}{2},n+\frac{N}{2},j} (A^*)_{n+p}^{\frac{N}{2},n+\frac{N}{2},j'} \quad (B.9)$$

needs to be evaluated. Use the expansion $(n + p + 1)^{1/2} \cong \sqrt{n+1} + \frac{p}{2\sqrt{n+1}}$. We also note that within the AFA that

$$(A^*)_{n+1}^{\frac{N}{2},n+\frac{N}{2}+1,j} A_{n+1}^{\frac{N}{2},n+\frac{N}{2}+1,j'} \cong (A^*)_n^{\frac{N}{2},n+\frac{N}{2},j} A_n^{\frac{N}{2},n+\frac{N}{2},j'}. \tag{B.10}$$

This means that we can use eq. B.4 to evaluate eq. B.9. Again, only the diagonal elements and elements with $j' = j \pm 1$ are non zero. For the off diagonal elements, eq.B.5 remains valid. For the diagonal elements it is found that

$$\sum_{p=0}^{N} p A_{n+p}^{\frac{N}{2},\left(n+\frac{N}{2}\right),j} (A^*)_{n+p}^{\frac{N}{2},\left(n+\frac{N}{2}\right),j} = \frac{N}{2} + \frac{\beta\left(j-\frac{N}{2}\right)}{2\sqrt{n_o}\sqrt{1+\frac{\bar{\beta}^2}{4}}}. \tag{B.11}$$

In order to proceed, we need to apply some approximations to the coefficients of time in Eqs. 21 or 26. Namely

$$q_{\frac{N}{2},n+\frac{N}{2},j} - q_{\frac{N}{2},n+\frac{N}{2}+1,j'} - \beta \cong -(r+n)\beta + (r-j)\sqrt{4+\beta^2} - (r+n+1)\beta + \left(r-j'\right)\sqrt{4+\beta^2} - \beta$$
$$\cong -(j-j')\sqrt{4+\beta^2} \tag{B.12}$$

The reason that this is an approximation is that the $\beta$ in eq. 26 is $\frac{\omega-\Omega}{\Omega|\kappa|}$ and when we switched to the AFA, there is another factor of $\sqrt{n_o}$ and $n_o$ increased by 1 when the good quantum number c is increased by 1. The approximation appears to be valid and of the same order as the approximation made for $(n+p+1)^{1/2}$. Using the above we can complete the expression for $S_2(\bar{n}, N, \gamma t)_{AFA}$.

$$S_2(\bar{n}, N, \gamma t)_{AFA} \cong \sum_{n=0}^{\infty} \langle n|\rho_f(0)|n+1\rangle$$
$$\times \left\{ \sum_{j=0}^{N} (A^*)_n^{\frac{N}{2},n+\frac{N}{2},j} A_n^{\frac{N}{2},n+\frac{N}{2},j} \left[(n+1)^{1/2} + \frac{N}{4(n+1)^{1/2}} + \frac{\beta\left(j-\frac{N}{2}\right)}{4\sqrt{n_o}(n+1)^{1/2}\sqrt{1+\frac{\bar{\beta}^2}{4}}}\right] \right.$$
$$\left. - \frac{N}{4(n+1)^{1/2}} \frac{n+\frac{N}{2}}{n+\frac{N}{2}+\Delta} Cos\left(\sqrt{4+\bar{\beta}^2}\sqrt{n_o}\Omega|\kappa|t\right)\right\} \tag{B.13}$$

The imaginary term goes to 0 in this approximation.

This reduces to

$$S_2(\bar{n}, N, \gamma t)_{AFA} \cong \sum_{n=0}^{\infty} \langle n|\rho_f(0)|n+1\rangle$$

$$\times \left\{ (n+1)^{1/2} + \frac{N}{4(n+1)^{1/2}} - \frac{N\beta^2}{16\left(n+\frac{N}{2}+\Delta\right)(n+1)^{1/2}} \right.$$

$$\left. - \frac{N}{4(n+1)^{1/2}} \frac{n+\frac{N}{2}}{n+\frac{N}{2}+\Delta} Cos\left(2\sqrt{n_0+\frac{\beta^2}{4}}\Omega|\kappa|t\right) \right\} \quad (B.14)$$

$$\cong \sum_{n=0}^{\infty} \langle n|\rho_f(0)|n+1\rangle \left\{ (n+1)^{1/2} + \frac{N}{2(n+1)^{1/2}} \frac{n+\frac{N}{2}}{n+\frac{N}{2}+\Delta} Sin^2\left(\sqrt{n+\frac{N}{2}+\Delta}\Omega|\kappa|t\right) \right\}$$

For resonance this reduces to

$$S_2(\bar{n}, N, \gamma t)_{AFA} \cong \sum_{n=0}^{\infty} \langle n|\rho_f(0)|n+1\rangle \left\{ (n+1)^{1/2} + \frac{N}{2(n+1)^{1/2}} Sin^2\left(\sqrt{n+\frac{N}{2}}\Omega|\kappa|t\right) \right\}, \Delta = 0. \quad (B.15)$$

As can be seen from the above discussion, the AFA leads to results much simpler and easy to understand. In addition, the term dependant on time decays as $\frac{1}{(n+1)^{1/2}}$. It would be interesting to evaluate the expressions to determine their validity.

## Some Numerical Justification

It would be interesting to determine under what conditions Eqs. B.8 and B.15 are valid. To determine validity, note that the initial equations for $S_1$ and $S_2$ (Eqs 24 and 26) depend on the differences of eigenvalues. For the approximation to be valid it is necessary that those differences be the same when separated by the same difference in eigenstate number. That is, all differences must be the same or close to the same for same differences in j. Thus if $j' = j \pm 1$ then all the $q_{\frac{N}{2}, n+\frac{N}{2}, j} - q_{\frac{N}{2}, n+\frac{N}{2}, j\pm 1}$ must have the same magnitude. The same is true for other $j, j'$ difference. For the $S_2$, it will be differences of $q_{\frac{N}{2}, n+\frac{N}{2}, j} - q_{\frac{N}{2}, n+\frac{N}{2}+1, j'}$ that will have to have the same behavior if the AFA is valid. To demonstrate the desired behavior, we plot several examples of these differences for resonance and non-resonances versus the number of photons. In Figure 1, we plot $q_{\frac{N}{2}, n+\frac{N}{2}, j} - q_{\frac{N}{2}, n+\frac{N}{2}, j'}$ for the 4 TLM case for both resonance and non-resonance.

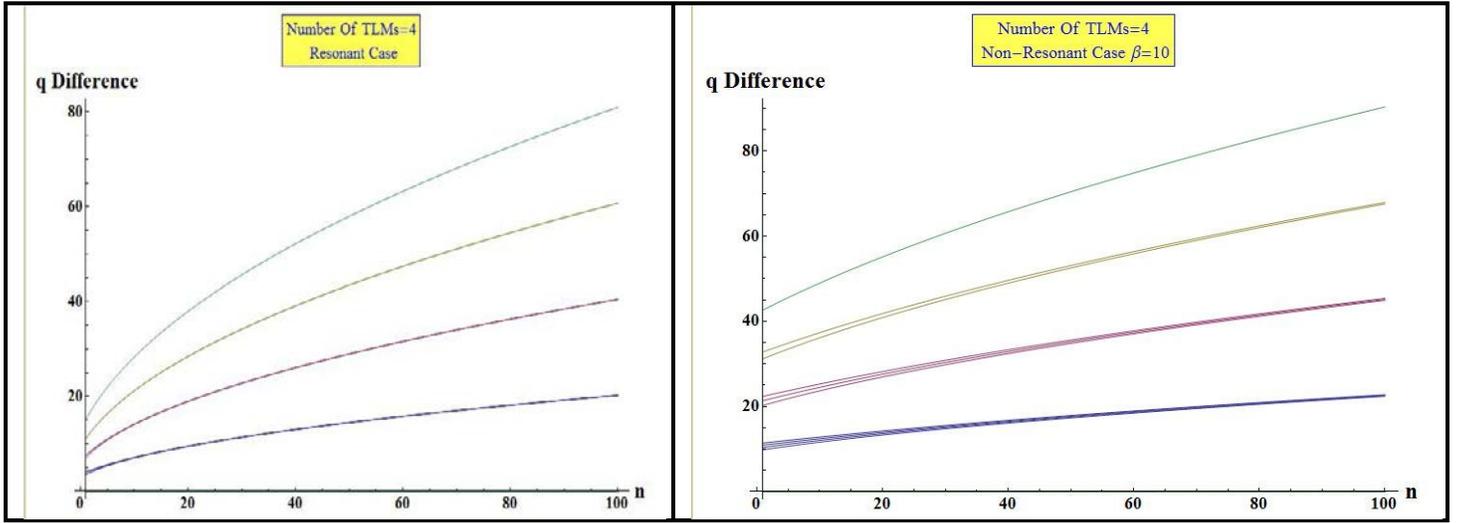

**Figure B1: q Differences for 4 TLMs. The lowest curve for j's differing by 1 while the upper curve the j's differ by 4.**

As can be seen, the q differences converge very quickly as n becomes large. For larger values of $\beta$ significantly larger values of n are required. A similar plot for $q_{\frac{N}{2},n+\frac{N}{2},j} - q_{\frac{N}{2},n+\frac{N}{2}+1,j'} - \beta$ is shown in the next figure.

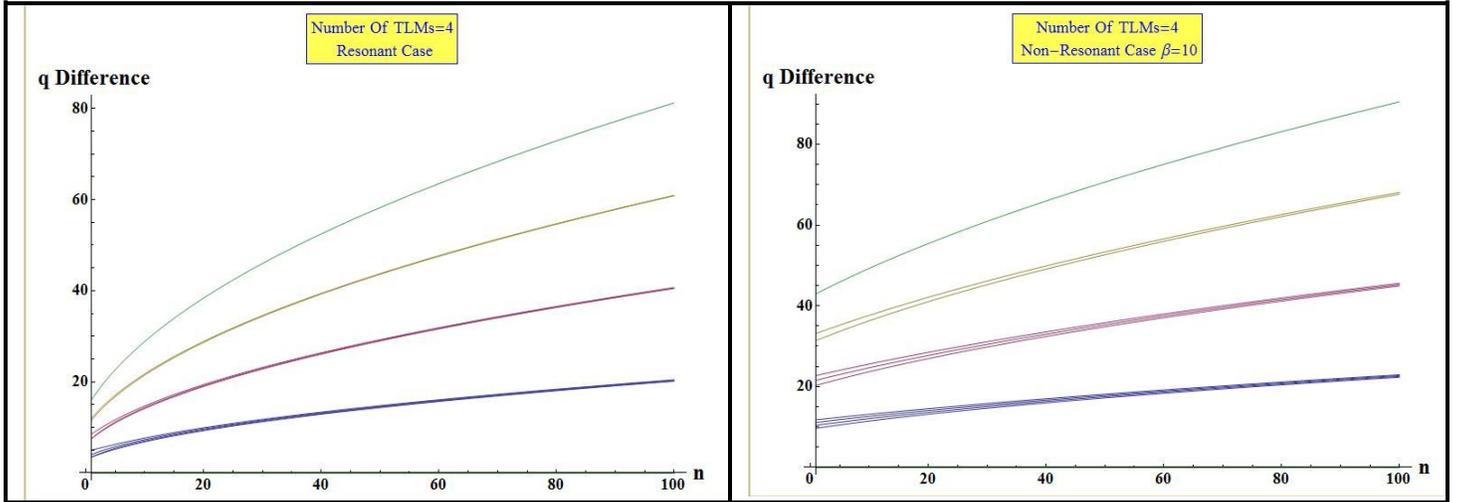

**Figure B2: q Differences used in the expression for $S_2$**

Figure B2 looks very similar to Figure B1. Closer inspection will show that the differences really never converge to the same value. There is always a difference of number less than unity between each of the sub-members of each set. However, on the scale of the magnitude of the differences, this difference becomes negligible as n becomes large.

With the verification that the trigonometric terms for the same j differences converge to the same value for large n, we examine the sum of the coefficients of the trigonometric terms for the same j differences. From eq. 24 we are then interested in the sum over j of the trigonometric terms with $j' = j + 1, j + 2, \cdots$. We want then

$$\sum_{j=1}^{N-k} (A^*)_n^{\frac{N}{2},\left(n+\frac{N}{2}\right),j} A_n^{\frac{N}{2},\left(n+\frac{N}{2}\right),j+k} \sum_{p=0}^{N} p A_{n+p}^{\frac{N}{2},\left(n+\frac{N}{2}\right),j} (A^*)_{n+p}^{\frac{N}{2},\left(n+\frac{N}{2}\right),j+k}, \quad k = 1,2,\cdots,N-1 \qquad (B.16)$$

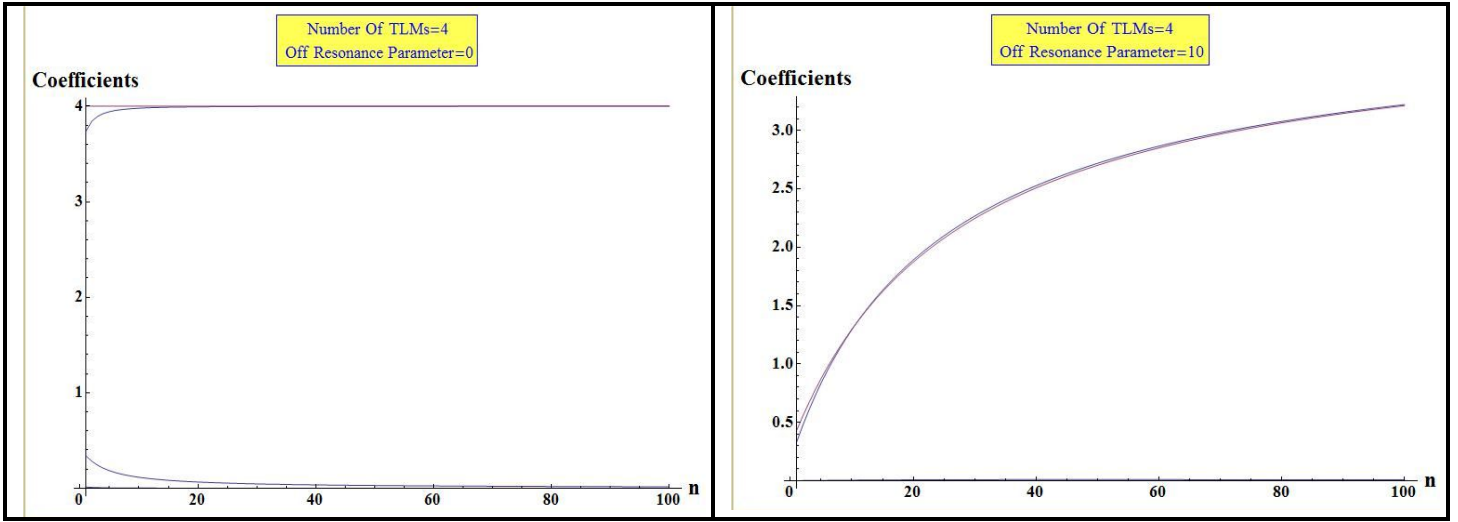

**Figure B3: Sum of the trigonometric term coefficients for the same $j, j'$ differences for $S_1$**

The top curve in Fig. B3 is the AFA coefficient seen in eq. B.8. The second curve is the sum of the coefficients with a j difference of 1. The other coefficients decay rapidly to 0 especially in the case of non-resonance. A similar set of coefficients for $S_2$ must be evaluated. Using eq. 26, the coefficients for the same j differences are given by

$$\sum_{j=0}^{N-k} (A^*)_{n+1}^{\frac{N}{2}, n+\frac{N}{2}+1, j} A_{n+1}^{\frac{N}{2}, n+\frac{N}{2}+1, j+k} \sum_{p=0}^{N} (n+p+1)^{1/2} A_{n+p}^{\frac{N}{2}, n+\frac{N}{2}, j} (A^*)_{n+p}^{\frac{N}{2}, n+\frac{N}{2}, j+k}, \quad -N \leq k \leq N. \tag{B.17}$$

We have performed the numerical evaluation of eq. B.17, for the case of $j' = j$ and also for the case of $j' = j \pm 1$ and compared them to the terms found in eq. B.14, namely to $(n+1)^{1/2} + \frac{N}{4(n+1)^{1/2}} \frac{n+\frac{N}{2}}{n+\frac{N}{2}+\Delta}$ for $j' = j$ and to

$-\frac{N}{4(n+1)^{1/2}} \frac{n+\frac{N}{2}}{n+\frac{N}{2}+\Delta}$ for $j' = j \pm 1$. The results are presented in Fig. B4 for N=10.

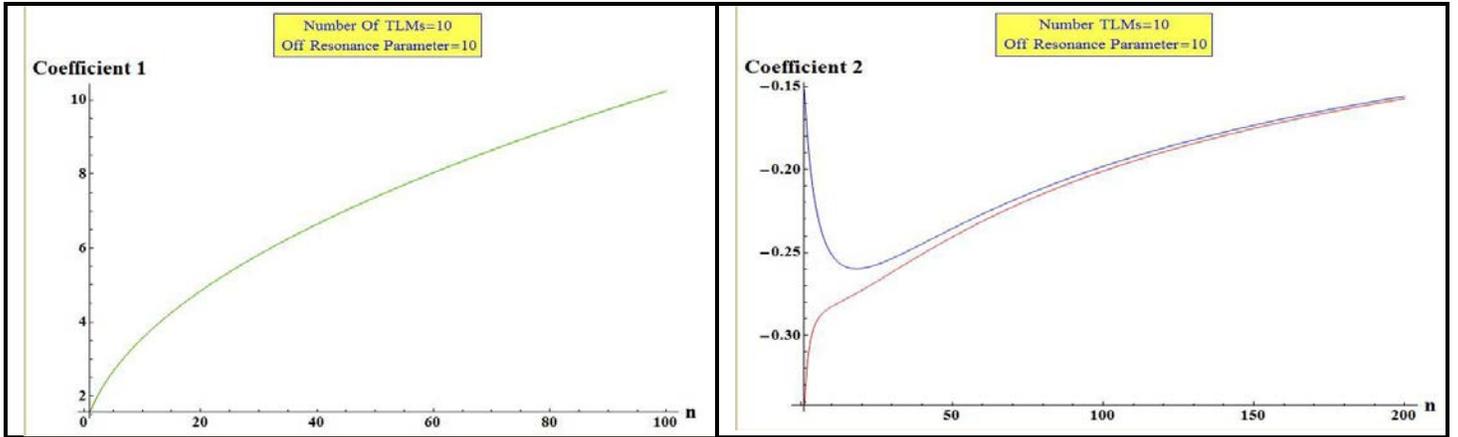

**Figure B4: Coefficients 1 and 2 from eq. 52 for $\beta = 10$.**

.In each of the figures above, the upper curve is from eq. B.17 and the bottom curve the approximation from eq. B.14 as discussed above. The curves on the left for $j' = j$ are nearly indistinguishable from each other while on the right the difference is easily explained as the error in the expansion of $(n+p+1)^{1/2}$ for smaller n. This completes the justification for the use of the AFA for large n.

# Appendix C: $S_1(\bar{n}, N, \gamma t)$ and $S_2(\bar{n}, N, \gamma t)$ for N=3 and 4 for Resonance.

## 3 TLMs

For N=3 and $\beta = 0$, $S_1(\bar{n}, 3, \gamma t)$ and $S_2(\bar{n}, 3, \gamma t)$ are given by;

$$
\begin{aligned}
S_1(\bar{n}, 3, \gamma t) = 4 \sum_{n=0}^{\infty} \langle n|\rho_f(0)|n\rangle \\
\times \Bigg\{ & \frac{3(2+n)(1+n+\sqrt{(1+n)(3+n)})\sin^2\left[\left(\sqrt{10+5n-\sqrt{73+16n(4+n)}} - \sqrt{10+5n+\sqrt{73+16n(4+n)}}\right)\frac{\Omega|\kappa|t}{2}\right]}{73+16n(4+n)} \\
+ & \frac{3(1+n)(2+n)(8+4n+\sqrt{73+16n(4+n)})\sin^2\left[\sqrt{10+5n-\sqrt{73+16n(4+n)}}\,\Omega|\kappa|t\right]}{2(73+16n(4+n))(-7-2n+\sqrt{73+16n(4+n)})} \\
- & \frac{3(2+n)(-1-n+\sqrt{(1+n)(3+n)})\sin^2\left[\left(\sqrt{10+5n-\sqrt{73+16n(4+n)}} + \sqrt{10+5n+\sqrt{73+16n(4+n)}}\right)\frac{\Omega|\kappa|t}{2}\right]}{73+16n(4+n)} \\
+ & \frac{3(1+n)(2+n)(-8-4n+\sqrt{73+16n(4+n)})\sin^2\left[\sqrt{10+5n+\sqrt{73+16n(4+n)}}\,\Omega|\kappa|t\right]}{2(73+16n(4+n))(7+2n+\sqrt{73+16n(4+n)})}\Bigg\}
\end{aligned}
\quad (C.1)
$$

$$
\begin{aligned}
S_2(\bar{n}, 3, \gamma t) = \sum_{n=0}^{\infty} & \frac{\langle n|\rho_f(0)|n+1\rangle}{4\sqrt{(73+64n+16n^2)(153+96n+16n^2)}} \\
\times \Bigg\{ & \left((7+2n)(\sqrt{1+n}-\sqrt{3+n}) + (\sqrt{1+n}+\sqrt{3+n})\sqrt{73+16n(4+n)}\right) \\
\times & \left(9+2n+\sqrt{153+16n(6+n)}\right) \cos\left[\sqrt{10+5n-\sqrt{73+16n(4+n)}}\,\Omega\kappa t\right] \cos\left[\sqrt{15+5n-\sqrt{73+16(1+n)(5+n)}}\,\Omega\kappa t\right] \\
- & 12(2+n)\left(-(1+n)\sqrt{3+n} + \sqrt{1+n}(3+n)\right) \cos\left[\sqrt{15+5n+\sqrt{153+96n+16n^2}}\,\Omega\kappa t\right] \cos\left[\sqrt{10+5n-\sqrt{73+16n(4+n)}}\,\Omega\kappa t\right] \\
+ & \left(-12(2+n)\left(-(1+n)\sqrt{3+n} + \sqrt{1+n}(3+n)\right) \cos\left[\sqrt{15+5n-\sqrt{153+96n+16n^2}}\,\Omega\kappa t\right]\right. \\
+ & \left((7+2n)(\sqrt{1+n}-\sqrt{3+n}) - (\sqrt{1+n}+\sqrt{3+n})\sqrt{73+16n(4+n)}\right) \\
\times & \left.\left(9+2n-\sqrt{153+16n(6+n)}\right) \cos\left[\sqrt{15+5n+\sqrt{153+96n+16n^2}}\,\Omega\kappa t\right]\right) \cos\left[\sqrt{10+5n+\sqrt{73+16n(4+n)}}\,\Omega\kappa t\right] \\
+ & ((1+2n)(-\sqrt{2+n}+\sqrt{4+n}) + (\sqrt{2+n}+\sqrt{4+n})\sqrt{73+16n(4+n)}) \\
\times & (9+2n+\sqrt{153+16n(6+n)})\sin[\sqrt{15+5n-\sqrt{153+96n+16n^2}}\,\Omega\kappa t]\sin[\sqrt{10+5n-\sqrt{73+16n(4+n)}}\,\Omega\kappa t] \\
+ & 12(2+n)(3+n)(\sqrt{2+n}-\sqrt{4+n})\sin[\sqrt{15+5n+\sqrt{153+96n+16n^2}}\,\Omega\kappa t]\sin[\sqrt{10+5n-\sqrt{73+16n(4+n)}}\,\Omega\kappa t] \\
+ & (12(2+n)(3+n)(\sqrt{2+n}-\sqrt{4+n})\sin[\sqrt{15+5n-\sqrt{153+96n+16n^2}}\,\Omega\kappa t] \\
+ & ((1+2n)(-\sqrt{2+n}+\sqrt{4+n}) - (\sqrt{2+n}+\sqrt{4+n})\sqrt{73+16n(4+n)}) \\
\times & (9+2n-\sqrt{153+16n(6+n)})\sin[\sqrt{15+5n+\sqrt{153+96n+16n^2}}\,\Omega\kappa t])\sin[\sqrt{10+5n+\sqrt{73+16n(4+n)}}\,\Omega\kappa t]\Bigg\}
\end{aligned}
\quad (C.2)
$$

## 4 TLMs

For N=4 and $\beta = 0$, $S_1(\bar{n}, 4, \gamma t)$ is given by;

$$S_1(\bar{n}, 4, \gamma t) = 4 \sum_{n=0}^{\infty} \langle n|\rho_f(0)|n\rangle$$

$$\times \left\{ \frac{(1+n)(2+n)(3+n)(10+4n+\sqrt{82+16n(5+n)})\mathrm{Sin}^2\left[\left(\sqrt{25+10n-3\sqrt{33+4n(5+n)}}-\sqrt{25+10n+3\sqrt{33+4n(5+n)}}\right)\frac{\Omega|\kappa|t}{2}\right]}{(33+4n(5+n))(41+8n(5+n))} \right.$$

$$+ \frac{12(1+n)(2+n)(3+n)(4+n)(1+2n+\sqrt{33+4n(5+n)})\mathrm{Sin}^2\left[\sqrt{25+10n-3\sqrt{33+4n(5+n)}}\frac{\Omega|\kappa|t}{2}\right]}{(41+8n(5+n))(561-87\sqrt{33+4n(5+n)}+n(505-45\sqrt{33+4n(5+n)}+2n(84+10n-3\sqrt{33+4n(5+n)}))))}$$

$$- \frac{12(-1-2n+\sqrt{33+4n(5+n)})\mathrm{Gamma}[5+n]\mathrm{Sin}^2\left[\sqrt{25+10+3\sqrt{33+4n(5+n)}}\frac{\Omega|\kappa|t}{2}\right]}{(41+8n(5+n))(561+87\sqrt{33+4n(5+n)}+n(505+20n^2+45\sqrt{33+4n(5+n)}+6n(28+\sqrt{33+4n(5+n)}))))n!}$$

$$- \frac{3(1+n)(2+n)(3+n)(9+\sqrt{33+4n(5+n)}+n(7+2n+\sqrt{33+4n(5+n)}))\mathrm{Sin}^2\left[\sqrt{25+10n-3\sqrt{33+4n(5+n)}}\Omega|\kappa|t\right]}{(561-87\sqrt{33+4n(5+n)}+n(505-45\sqrt{33+4n(5+n)}+2n(84+10n-3\sqrt{33+4n(5+n)})))^2}$$

$$- \frac{(1+n)(2+n)(3+n)(-10-4n+\sqrt{82+16n(5+n)})\mathrm{Sin}^2\left[\left(\sqrt{25+10n-3\sqrt{33+4n(5+n)}}+\sqrt{25+10n+3\sqrt{33+4n(5+n)}}\right)\frac{\Omega|\kappa|t}{2}\right]}{(33+4n(5+n))(41+8n(5+n))}$$

$$+ \left. \frac{3(1+n)(2+n)(3+n)(-9+\sqrt{33+4n(5+n)}+n(-7-2n+\sqrt{33+4n(5+n)}))\mathrm{Sin}^2\left[\sqrt{25+10n+3\sqrt{33+4n(5+n)}}\Omega|\kappa|t\right]}{(561+87\sqrt{33+4n(5+n)}+n(505+20n^2+45\sqrt{33+4n(5+n)}+6n(28+\sqrt{33+4n(5+n)})))^2} \right\} \quad (C.3)$$

The expression for $S_2(\bar{n}, 4, \gamma t)$ is not provided due to its complexity.

## Appendix D: Greater than one TLMs Non Resonant

For N≥1, the analytic expressions for $S_1(\bar{n}, N, \gamma t)_{NR}$ and $S_2(\bar{n}, N, \gamma t)_{NR}$ become completely unwieldy (intractable), for instance the eigenvectors are found by substituting the eigenvalues into the general expression for the eigenvector for the 2 TLM case is given by

$$b = \left\{ \begin{array}{c} 2\sqrt{\frac{1}{4+\frac{2(q+n\beta)^2}{n+1}+\frac{[-2+q(q+\beta)+n(-2+\beta\{2q+\beta(n+1)\})]^2}{(n+1)(n+2)}}} \\ \sqrt{\frac{2(q+n\beta)^2}{(n+1)\left[4+\frac{2(q+n\beta)^2}{n+1}+\frac{[-2+q(q+\beta)+n(-2+\beta\{2q+\beta(n+1)\})]^2}{(n+1)(n+2)}\right]}} \\ \sqrt{\frac{\{-2+q(q+\beta)+n[-2+\beta\{2q+\beta(n+1)\}]\}^2}{(n+1)\left[4+\frac{2(q+n\beta)^2}{n+1}+\frac{[-2+q(q+\beta)+n(-2+\beta\{2q+\beta(n+1)\})]^2}{(n+1)(n+2)}\right]}} \end{array} \right\} \quad (D.1)$$

The effective eigenvalues are given by

$$q_j = \begin{Bmatrix} -\beta - n\beta + \dfrac{(2)^{\frac{1}{3}}(6 + 2n + \beta^2)}{[-54\beta + \sqrt{2916\beta^2 - 108(6 + 2n + \beta^2)^3}]^{\frac{1}{3}}} + \dfrac{[-54\beta + \sqrt{2916\beta^2 - 108(6 + 2n + \beta^2)^3}]^{\frac{1}{3}}}{3(2)^{\frac{1}{3}}} \\ -\beta - n\beta - \dfrac{(1 - \sqrt{3}i)(6 + 2n + \beta^2)}{(2)^{\frac{2}{3}}[-54\beta + \sqrt{2916\beta^2 - 108(6 + 2n + \beta^2)^3}]^{\frac{1}{3}}} - \dfrac{(1 + \sqrt{3}i)[-54\beta + \sqrt{2916\beta^2 - 108(6 + 2n + \beta^2)^3}]^{\frac{1}{3}}}{6(2)^{\frac{1}{3}}} \\ -\beta - n\beta - \dfrac{(1 + \sqrt{3}i)(6 + 2n + \beta^2)}{(2)^{\frac{2}{3}}[-54\beta + \sqrt{2916\beta^2 - 108(6 + 2n + \beta^2)^3}]^{\frac{1}{3}}} - \dfrac{(1 - \sqrt{3}i)[-54\beta + \sqrt{2916\beta^2 - 108(6 + 2n + \beta^2)^3}]^{\frac{1}{3}}}{6(2)^{\frac{1}{3}}} \end{Bmatrix} \quad (D.2)$$

It is obvious that these expressions can be simplified somewhat by pulling out the common factor of 3 from the cube roots; however even with this simplification, when the q's are substituted into eq. D.1, the resulting eigenvectors and thus the factors needed to express $S_1$ and $S_2$ in terms of n and $\beta$ become exceedingly complex. Note that even though eq. D.2 appears to provide complex results for the eigenvalues, they are in fact real. Numerical calculation for eq. D.2 would find that the complex components for the eigenvalues are essentially 0 to machine accuracy. On the other hand performing a complex expansion of eq. D.2 before evaluation gives exactly 0 for the complex part.

For example performing the complex expansion on eq. D.2, one obtains

$$q_j = \begin{Bmatrix} -\beta - n\beta + \dfrac{2\sqrt{y2}\cos[\frac{1}{3}\text{Arg}[z3]]}{\sqrt{3}} \\ -\beta - n\beta - \dfrac{1}{3}\sqrt{y2}(\sqrt{3}\cos[\frac{1}{3}\text{Arg}[z3]] - 3\sin[\frac{1}{3}\text{Arg}[z3]]) \\ -\beta - n\beta - \dfrac{1}{3}\sqrt{y2}(\sqrt{3}\cos[\frac{1}{3}\text{Arg}[z3]] + 3\sin[\frac{1}{3}\text{Arg}[z3]]) \end{Bmatrix} \quad (D.3)$$

where y2=6+2n+$\beta^2$ and z3=-54$\beta$+$\sqrt{2916\beta^2 - 108(y2)^3}$. These are obviously real expressions. The Arg function yields the phase angle of z3 with a value between $-\pi$ and $\pi$. Note that, since the imaginary part of z3 is always positive and greater than 0, then the phase angle is centered about $\frac{\pi}{2}$.

# Appendix E: Spontaneous Emission

## All TLMs Initially Up

It is of interest to consider various cases of spontaneous emission. In order to do so, consider first the case when there are no photons in the field at time 0 and all TLMs are in the up state. The following table contains the analytic results for 1, 2 and 3 TLMs as well as the numerical results for 50 TLMs all for the resonant case. The first 3 results are based on Eqs. (29b), (30a), (C.1).

$$S_1(0,1,\gamma t) = Sin^2[\Omega|\kappa|t] : N = 1, \beta = 0.$$

$$S_1(0,2,\gamma t) = 8\left\{\frac{2}{9}Sin^2\left[\sqrt{\frac{3}{2}}\Omega|\kappa|t\right] - \frac{1}{72}Sin^2\left[2\sqrt{\frac{3}{2}}\Omega|\kappa|t\right]\right\}$$

$$S_1(0,3,\gamma t) = \frac{12}{73}\left\{2(1+\sqrt{3})Sin^2\left[\left(\sqrt{10-\sqrt{73}} - \sqrt{10+\sqrt{73}}\right)\frac{\Omega|\kappa|t}{2}\right]\right.$$
$$- 2(-1+\sqrt{3})Sin^2\left[\left(\sqrt{10-\sqrt{73}} + \sqrt{10+\sqrt{73}}\right)\frac{\Omega|\kappa|t}{2}\right]$$
$$\left.+ \frac{(8+\sqrt{73})Sin^2\left[\sqrt{10-\sqrt{73}}\Omega|\kappa|t\right]}{(-7+\sqrt{73})} + \frac{(-8+\sqrt{73})Sin^2\left[\sqrt{10+\sqrt{73}}\Omega|\kappa|t\right]}{(7+\sqrt{73})}\right\}$$

(E.1)

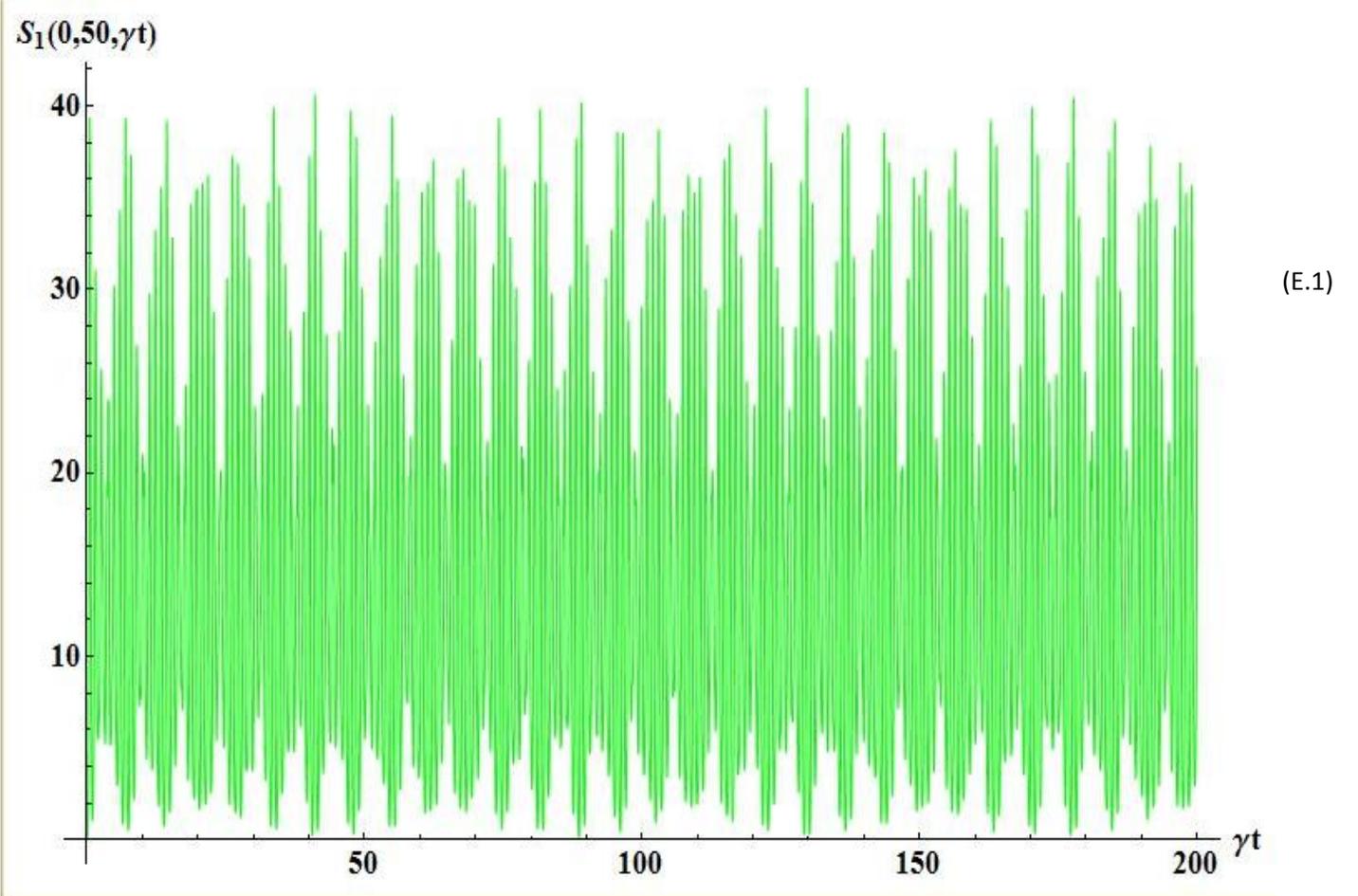

**Figure E 1:** Numerical Solution for $S_1(0, 50, \gamma t)$ for the resonant case all TLMs initially up.

Examination of E.1 as time goes to zero shows $S_1(0, N, \gamma t \to 0)/\gamma t^2 = N$. In fact these results can be found analytically. If one starts from eq. (23) and specifies that at time 0, that there are no photons in the field[!], then

$$S_1(0, N, \gamma t) = \sum_{j=0}^{N} \sum_{j'=0}^{N} e^{-i\left(\lambda_{\frac{N}{2},\frac{N}{2},j} - \lambda_{\frac{N}{2},\frac{N}{2},j'}\right)t} (A^*)_0^{\frac{N}{2},\frac{N}{2},j} A_0^{\frac{N}{2},\frac{N}{2},j'}$$
$$\times \sum_{p=0}^{N} p A_p^{\frac{N}{2},\frac{N}{2},j} (A^*)_p^{\frac{N}{2},\frac{N}{2},j'}$$
(E.2)

Expand the exponential out to the squared term in time yielding

$$S_1(0, N, \gamma t) = \sum_{j=0}^{N} \sum_{j'=0}^{N} \left[1 - i\left(\lambda_{\frac{N}{2},\frac{N}{2},j} - \lambda_{\frac{N}{2},\frac{N}{2},j'}\right)t - \left(\lambda_{\frac{N}{2},\frac{N}{2},j} - \lambda_{\frac{N}{2},\frac{N}{2},j'}\right)^2 \frac{t^2}{2} \cdots\right] (A^*)_0^{\frac{N}{2},\frac{N}{2},j} A_0^{\frac{N}{2},\frac{N}{2},j'}$$
$$\times \sum_{p=0}^{N} p A_p^{\frac{N}{2},\frac{N}{2},j} (A^*)_p^{\frac{N}{2},\frac{N}{2},j'}$$
E.3

The first two terms are identically 0, since for the constant term the only existing factor is for p=0 and for the second, the difference is 0 since you can replace j by j'. This leaves the third term.

$$S_1(0, N, \gamma t) = -\frac{t^2}{2} \sum_{j=0}^{N} \sum_{j'=0}^{N} \left(\lambda_{\frac{N}{2},\frac{N}{2},j} - \lambda_{\frac{N}{2},\frac{N}{2},j'}\right)^2 (A^*)_0^{\frac{N}{2},\frac{N}{2},j} A_0^{\frac{N}{2},\frac{N}{2},j'} \sum_{p=0}^{N} p A_p^{\frac{N}{2},\frac{N}{2},j} (A^*)_p^{\frac{N}{2},\frac{N}{2},j'}$$
$$= -\frac{(\Omega|\kappa|t)^2}{2} \sum_{j=0}^{N} \sum_{j'=0}^{N} \left(q^2_{\frac{N}{2},\frac{N}{2},j} + q^2_{\frac{N}{2},\frac{N}{2},j'} - 2q_{\frac{N}{2},\frac{N}{2},j}q_{\frac{N}{2},\frac{N}{2},j'}\right) (A^*)_0^{\frac{N}{2},\frac{N}{2},j} A_0^{\frac{N}{2},\frac{N}{2},j'} \sum_{p=0}^{N} p A_p^{\frac{N}{2},\frac{N}{2},j} (A^*)_p^{\frac{N}{2},\frac{N}{2},j'}$$
$$= -(\Omega|\kappa|t)^2 \left\{\sum_{p=0}^{N} p \left[\sum_{j=0}^{N} q^2_{\frac{N}{2},\frac{N}{2},j} (A^*)_0^{\frac{N}{2},\frac{N}{2},j} A_p^{\frac{N}{2},\frac{N}{2},j} \sum_{j'=0}^{N} A_0^{\frac{N}{2},\frac{N}{2},j'} (A^*)_p^{\frac{N}{2},\frac{N}{2},j'} - \left|\sum_{j=0}^{N} q_{\frac{N}{2},\frac{N}{2},j} (A^*)_0^{\frac{N}{2},\frac{N}{2},j} A_p^{\frac{N}{2},\frac{N}{2},j}\right|^2\right]\right\}$$
(E.4)

I have replaced $\lambda_{\frac{N}{2},\frac{N}{2},j}$ by the effective Eigenvalue and made the appropriate change in time variable. The only factor for the first term that exits is for p=0 due to orthogonality. Further an examination of the Eigenvectors shows that

$$q_{\frac{N}{2},\frac{N}{2},j} (A^*)_0^{\frac{N}{2},\frac{N}{2},j} = \sqrt{N} (A^*)_1^{\frac{N}{2},\frac{N}{2},j}$$
(E.5)

Again, due to orthogonality the only existing factor for the second term is for p=1 with the results

$$S_1(0, N, \gamma t) = (\Omega|\kappa|t)^2 \left|\sum_{j=0}^{N} q_{\frac{N}{2},\frac{N}{2},j} (A^*)_0^{\frac{N}{2},\frac{N}{2},j} A_1^{\frac{N}{2},\frac{N}{2},j}\right|^2 = N(\Omega|\kappa|t)^2, QED$$
(E.6)

The results indicate that super-radiant behavior[9] is not seen for this case which is not unexpected since $m = r = \frac{N}{2}$.

---

[!] eq. F.15 of Reference 6, contains an error in that the factor representing the number of photons was not handled properly.

## Only One or Two TLMs Up.

A second case of interest would be for all TLMs down except for 1 or 2 TLMs which are in the up state. Again we assume the resonant condition due to the ability to display relatively simple analytic results.

Start with eq. (18), and assume that there are no photons initially in the field. Further for the first example, assume that all TLMs except 1 are in the lower state. That implies $p = 1 - \frac{N}{2}$. Then there are two choices for r, namely $\frac{N}{2}$ or $\frac{N}{2} - 1$. The choice for $r = \frac{N}{2} - 1$ is not viable since c-m will be zero. This leaves the results that

$$\langle E^- E^+(t) \rangle = \left|\frac{\gamma}{\mu}\right|^2 Sin^2(\sqrt{N}\gamma t) \langle \frac{N}{2}, 1 - \frac{N}{2} | \rho_{TLM}(0) | \frac{N}{2}, 1 - \frac{N}{2} \rangle \tag{E.7}$$

To complete the analysis it is necessary to determine the density matrix at time 0. One example would be for a given TLM to be in the up state. It doesn't matter which one. For instance

$$\rho_{TLM}(0) = |-----\cdots +\rangle\langle -----\cdots +| \tag{E.8}$$

Then

$$\langle \frac{N}{2}, 1 - \frac{N}{2} | \rho_{TLM}(0) | \frac{N}{2}, 1 - \frac{N}{2} \rangle = \frac{1}{N}. \tag{E.9}$$

For that case we find that

$$\langle E^- E^+(t) \rangle = \left|\frac{\gamma}{\mu}\right|^2 \frac{Sin^2(\sqrt{N}\gamma t)}{N} \tag{E.10}$$

Another case is that we don't know which TLM is up with all TLMs being equally likely to be up. For that case the density matrix of the TLMs is given by

$$\rho_{TLM}(0) = |\frac{N}{2}, 1 - \frac{N}{2}\rangle\langle\frac{N}{2}, 1 - \frac{N}{2}| \tag{E.11}$$

$$\langle \frac{N}{2}, 1 - \frac{N}{2} | \rho_{TLM}(0) | \frac{N}{2}, 1 - \frac{N}{2} \rangle = 1.. \tag{E.12}$$

The ensemble average of the number operator is given by

$$\langle E^- E^+(t) \rangle = \left|\frac{\gamma}{\mu}\right|^2 Sin^2(\sqrt{N}\gamma t) \tag{E.13}$$

Thus we observe damping of the number density when an initial state for the TLMs is specified.[10] Both cases were discussed in the reference.

---

[9] R. H. Dicke, Coherence in Spontaneous Radiation Processes, Phy. Rev. Vol. 93, #1, Jan. 1, 1954
[10] F. W. Cummings and Ali Dorri, Exact Solution for Spontaneous Emission in the presence on N Atoms, Phys. Rev. A, Vol. 28, #4, Oct. 1983

As a second example, consider all TLMs except 2 in the down state. Following the same procedure for the single case

$$\langle E^- E^+(t) \rangle = \left|\frac{\gamma}{\mu}\right|^2 \sum_{r=0,\frac{1}{2}}^{\frac{N}{2}} P(r) \sum_{m=-r}^{r} \left(2 - \frac{N}{2} - m\right) \sum_{j=0}^{2-\frac{N}{2}+r} \sum_{j'=0}^{2-\frac{N}{2}+r} (A^*)_0^{r,2-\frac{N}{2},j}$$
$$\times A_{2-\frac{N}{2}-m}^{r,2-\frac{N}{2},j} (A^*)_{2-\frac{N}{2}-m}^{r,2-\frac{N}{2},j'} A_0^{r,2-\frac{N}{2},j'} e^{-i\lambda_{r,2-\frac{N}{2},j} t} e^{i\lambda_{r,2-\frac{N}{2},j'} t} \langle r, 2 - \frac{N}{2} | \rho_{TLM}(0) | r, 2 - \frac{N}{2} \rangle$$
(E.14)

Valid values or r are $\frac{N}{2}, \frac{N}{2} - 1$ and $\frac{N}{2} - 2$. The last value results in $\left(2 - \frac{N}{2} - m\right) = 0$, thus only the first two values of r are provide non zero results. For r= $\frac{N}{2} - 1$, there is a degeneracy factor of N-1. However each of the factors has exactly the same time dependence of $Sin^2(\sqrt{N-1}\gamma t)$ based on the 1 TLM analyses discussed above. The term for r=$\frac{N}{2}$ is given by

$$\langle E^- E^+(t) \rangle = \left|\frac{\gamma}{\mu}\right|^2 \sum_{m=-\frac{N}{2}}^{-\frac{N}{2}+1} \left(2 - \frac{N}{2} - m\right) \sum_{j=0}^{2} \sum_{j'=0}^{2} (A^*)_0^{\frac{N}{2},2-\frac{N}{2},j}$$
$$\times A_{2-\frac{N}{2}-m}^{\frac{N}{2},2-\frac{N}{2},j} (A^*)_{2-\frac{N}{2}-m}^{\frac{N}{2},2-\frac{N}{2},j'} A_0^{\frac{N}{2},2-\frac{N}{2},j'} e^{-i\lambda_{\frac{N}{2},2-\frac{N}{2},j} t} e^{i\lambda_{\frac{N}{2},2-\frac{N}{2},j'} t} \langle \frac{N}{2}, 2 - \frac{N}{2} | \rho_{TLM}(0) | \frac{N}{2}, 2 - \frac{N}{2} \rangle$$
$$= \left|\frac{\gamma}{\mu}\right|^2 \langle \frac{N}{2}, 2 - \frac{N}{2} | \rho_{TLM}(0) | \frac{N}{2}, 2 - \frac{N}{2} \rangle \sum_{j=0}^{2} (A^*)_0^{\frac{N}{2},2-\frac{N}{2},j} e^{-i\lambda_{\frac{N}{2},2-\frac{N}{2},j} t} \sum_{j'=0}^{2} e^{-i\lambda_{\frac{N}{2},2-\frac{N}{2},j'} t} A_0^{\frac{N}{2},2-\frac{N}{2},j'} \sum_{p=0}^{1} (2-p) A_{2-p}^{\frac{N}{2},2-\frac{N}{2},j} (A^*)_{2-p}^{\frac{N}{2},2-\frac{N}{2},j'},$$
(E.15)

by a change of variables and reordering terms.

Expanding terms in eq. (E.15),

$$\langle E^- E^+(t) \rangle = \left|\frac{\gamma}{\mu}\right|^2 \langle \frac{N}{2}, 2 - \frac{N}{2} | \rho_{TLM}(0) | \frac{N}{2}, 2 - \frac{N}{2} \rangle \left[ 2\left|\sum_{j=0}^{2} (A^*)_0^{\frac{N}{2},2-\frac{N}{2},j} A_2^{\frac{N}{2},2-\frac{N}{2},j} e^{-i\lambda_{\frac{N}{2},2-\frac{N}{2},j} t}\right|^2 + \left|\sum_{j=0}^{2} (A^*)_0^{\frac{N}{2},2-\frac{N}{2},j} A_1^{\frac{N}{2},2-\frac{N}{2},j} e^{-i\lambda_{\frac{N}{2},2-\frac{N}{2},j} t}\right|^2 \right]$$
$$= \left|\frac{\gamma}{\mu}\right|^2 \langle \frac{N}{2}, 2 - \frac{N}{2} | \rho_{TLM}(0) | \frac{N}{2}, 2 - \frac{N}{2} \rangle \left\{ \frac{4N(N-1)}{(2N-1)^2} Sin^4 \left[\frac{\sqrt{2N-1}\gamma t}{\sqrt{2}}\right] + \frac{N-1}{2N-1} Sin^2 \left[\sqrt{2}\sqrt{2N-1}\gamma t\right] \right\}$$
(E.16)
$$= \left|\frac{\gamma}{\mu}\right|^2 \langle \frac{N}{2}, 2 - \frac{N}{2} | \rho_{TLM}(0) | \frac{N}{2}, 2 - \frac{N}{2} \rangle \left[ \frac{4(N-1)}{2N-1} Sin^2 \left[\frac{\sqrt{2N-1}\gamma t}{\sqrt{2}}\right] \right] \left[ 1 + \frac{1}{2N-1} Sin^2 \left[\frac{\sqrt{2N-1}\gamma t}{\sqrt{2}}\right] \right] \quad (term\ 1)$$

To this is added the N-1 terms, each of which looks like

$$\langle E^- E^+(t) \rangle = \left|\frac{\gamma}{\mu}\right|^2 \langle \frac{N}{2} - 1, 2 - \frac{N}{2} | \rho_{TLM}(0) | \frac{N}{2} - 1, 2 - \frac{N}{2} \rangle Sin^2(\sqrt{N-1}\gamma t)$$
(E.17)

If one assumes that the 2 up TLMs are specified and it doesn't matter which ones then

$$\langle \frac{N}{2}, 2 - \frac{N}{2} | \rho_{TLM}(0) | \frac{N}{2}, 2 - \frac{N}{2} \rangle = \frac{2}{N(N-1)}$$
(E.18)

$$\sum_{(N-1) terms} \langle \frac{N}{2} - 1, 2 - \frac{N}{2} | \rho_{TLM}(0) | \frac{N}{2} - 1, 2 - \frac{N}{2} \rangle = \frac{2}{N}$$

The final results is

$$\langle E^- E^+(t) \rangle = \left|\frac{\gamma}{\mu}\right|^2 \left\{ \frac{8}{N(2N-1)} Sin^2\left[\frac{\sqrt{2N-1}\gamma t}{\sqrt{2}}\right]\left[1 + \frac{1}{2N-1} Sin^2\left[\frac{\sqrt{2N-1}\gamma t}{\sqrt{2}}\right]\right] + \frac{2}{N} Sin^2(\sqrt{N-1}\gamma t) \right\} \quad (E.19)$$

This again shows the strong damping to spontaneous emission exerted by the other down state TLMs.

## Same Number of TLMs Up as Down

Assume that you have an initial state with 0 photons, r=$\frac{N}{2}$ and m=0. That is you are in the Dicke potential radiant state. Then c=0 and the field density is given by

$$\langle E^- E^+(t) \rangle = \left|\frac{\gamma}{\mu}\right|^2 \sum_{p=1}^{\frac{N}{2}} p \sum_{j=0}^{\frac{N}{2}} \sum_{j'=0}^{\frac{N}{2}} e^{-i\left(\lambda_{\frac{N}{2},0,j} - \lambda_{\frac{N}{2},0,j'}\right)t} (A^*)_0^{\frac{N}{2},0,j} A_p^{\frac{N}{2},0,j} (A^*)_0^{\frac{N}{2},0,j'} A_p^{\frac{N}{2},0,j'} \quad (E.20)$$

This can also be written as

$$\langle E^- E^+(t) \rangle = -4 \left|\frac{\gamma}{\mu}\right|^2 \sum_{j=0}^{\frac{N}{2}-1} \sum_{j'=j+1}^{\frac{N}{2}} Sin^2\left\{\frac{\left[q_{\frac{N}{2},0,j} - q_{\frac{N}{2},0,j'}\right]}{2}\Omega|\kappa|t\right\} (A^*)_0^{\frac{N}{2},0,j} A_0^{\frac{N}{2},0,j'} \sum_{p=0}^{\frac{N}{2}} p A_p^{\frac{N}{2},0,j} (A^*)_p^{\frac{N}{2},0,j'} \quad (E.21)$$

Note that the upper limits really depend on N being even. There is a minor change for odd N. These two results look remarkably like the case for all TLMs in the up state but the detailed results is considerably different. For instance, if you follow the same analysis for the time behavior near 0 as used for the all up TLM case Eqs. E.4 through E.6, the results is that

$$S_1(0, N, \gamma t) = (\Omega|\kappa|t)^2 \left|\sum_{j=0}^{\frac{N}{2}} q_{\frac{N}{2},0,j} (A^*)_0^{\frac{N}{2},0,j} A_1^{\frac{N}{2},0,j}\right|^2 = \left(\frac{N}{2}\right)^2 (\Omega|\kappa|t)^2, for\ m = 0 \quad (E.22)$$

This demonstrates Super Radiance and is consistent with the Dicke formulation for m=0. In addition it is found that all the energy stored in the TLMs is dumped to the field in a very short time. Finally, the longer time behavior is demonstrated and shown to be very different than the all up TLM case. In fact we see what appears to decay and revival of oscillation in the following figure.

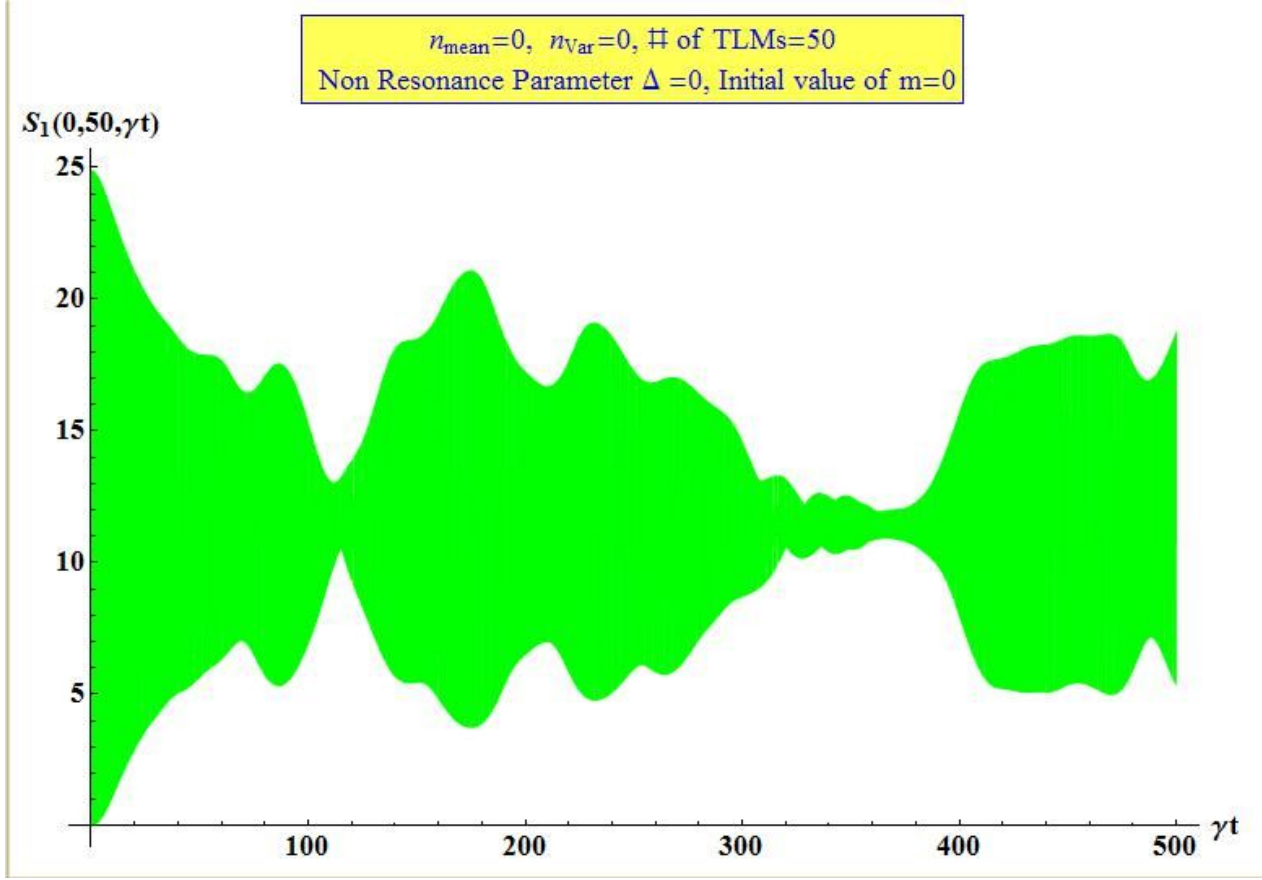

Figure E 2: Numerical Solution for $S_1(0, 50, \gamma t)$ for the resonant case half of TLMs initially up.

# Appendix F: Stimulated Absorption

We use the same formulation to examine stimulated absorption by taking the initial TLM state to be with all TLMs in the down state, that is that $m=-\frac{N}{2}$. Starting with eq. (18)

$$\langle E^-E^+(t)\rangle = \left|\frac{\gamma}{\mu}\right|^2 \sum_{m=-\frac{N}{2}}^{\frac{N}{2}} \sum_{c=m}^{\infty} (c-m) \sum_{j=0}^{Min[N,c+\frac{N}{2}]} \sum_{j'=0}^{Min[N,c+\frac{N}{2}]} (A^*)^{\frac{N}{2},c,j}_{c+\frac{N}{2}} \times A^{\frac{N}{2},c,j}_{c-m} (A^*)^{\frac{N}{2},c,j'}_{c-m} A^{\frac{N}{2},c,j'}_{c+\frac{N}{2}} e^{-i\lambda_{\frac{N}{2},c,j}t} e^{i\lambda_{\frac{N}{2},c,j'}t} \langle c+\frac{N}{2}|\rho_f(0)|c+\frac{N}{2}\rangle \quad (F.1)$$

Let $p=m+\frac{N}{2}$, then

$$\langle E^-E^+(t)\rangle = \left|\frac{\gamma}{\mu}\right|^2 \sum_{p=0}^{N} \sum_{c=p-\frac{N}{2}}^{\infty} \left(c+\frac{N}{2}-p\right) \sum_{j=0}^{Min[N,c+\frac{N}{2}]} \sum_{j'=0}^{Min[N,c+\frac{N}{2}]} (A^*)^{\frac{N}{2},c,j}_{c+\frac{N}{2}} \times A^{\frac{N}{2},c,j}_{c+\frac{N}{2}-p} (A^*)^{\frac{N}{2},c,j'}_{c+\frac{N}{2}-p} A^{\frac{N}{2},c,j'}_{c+\frac{N}{2}} e^{-i\lambda_{\frac{N}{2},c,j}t} e^{i\lambda_{\frac{N}{2},c,j'}t} \langle c+\frac{N}{2}|\rho_f(0)|c+\frac{N}{2}\rangle \quad (F.2)$$

Let $n=c+\frac{N}{2}$, yielding

$$\langle E^-E^+(t)\rangle = \left|\frac{\gamma}{\mu}\right|^2 \sum_{p=0}^{N} \sum_{n=p}^{\infty} (n-p) \sum_{j=0}^{Min[N,n]} \sum_{j'=0}^{Min[N,n]} (A^*)^{\frac{N}{2},n-\frac{N}{2},j}_{n} \times A^{\frac{N}{2},n-\frac{N}{2},j}_{n-p} (A^*)^{\frac{N}{2},n-\frac{N}{2},j'}_{n-p} A^{\frac{N}{2},n-\frac{N}{2},j'}_{n} e^{-i\lambda_{\frac{N}{2},n-\frac{N}{2},j}t} e^{i\lambda_{\frac{N}{2},n-\frac{N}{2},j'}t} \langle n|\rho_f(0)|n\rangle \quad (F.3)$$

Reorder the summation over n and p

$$\langle E^-E^+(t)\rangle = \left|\frac{\gamma}{\mu}\right|^2 \sum_{n=0}^{\infty} \sum_{p=0}^{\min[N,n]} (n-p) \sum_{j=0}^{Min[N,n]} \sum_{j'=0}^{Min[N,n]} (A^*)_n^{\frac{N}{2},n-\frac{N}{2},j} A_{n-p}^{\frac{N}{2},n-\frac{N}{2},j} (A^*)_{n-p}^{\frac{N}{2},n-\frac{N}{2},j'} A_n^{\frac{N}{2},n-\frac{N}{2},j'} e^{-i\lambda_{\frac{N}{2},n-\frac{N}{2},j}t} e^{i\lambda_{\frac{N}{2},n-\frac{N}{2},j'}t} \langle n|\rho_f(0)|n\rangle \qquad (F.4)$$

Using the same orthogonality relationships used previously

$$\langle E^-E^+(t)\rangle = \left|\frac{\gamma}{\mu}\right|^2 \left\{\bar{n} - \sum_{n=0}^{\infty}\langle n|\rho_f(0)|n\rangle \sum_{j=0}^{Min[N,n]}\sum_{j'=0}^{Min[N,n]} (A^*)_n^{\frac{N}{2},n-\frac{N}{2},j} A_n^{\frac{N}{2},n-\frac{N}{2},j'} e^{i\lambda_{\frac{N}{2},n-\frac{N}{2},j}t} e^{-i\lambda_{\frac{N}{2},n-\frac{N}{2},j'}t} \sum_{p=0}^{min[N,n]} pA_{n-p}^{\frac{N}{2},n-\frac{N}{2},j} (A^*)_{n-p}^{\frac{N}{2},n-\frac{N}{2},j'}\right\} \qquad (F.5)$$

This expression is very similar to Eqs. (22)and (23) but with some differences since the upper limits on the sums over j, j' and p depend on n and the Eigenvectors $A_k^{r,c,j}$ are summed downward from the highest term rather than upwards from the lowest term. As for the all up TLM case we consider further only the second term in (F.5).

$$S_4(\bar{n}, N, \gamma t) = \sum_{n=0}^{\infty}\langle n|\rho_f(0)|n\rangle \sum_{j=0}^{Min[N,n]}\sum_{j'=0}^{Min[N,n]} (A^*)_n^{\frac{N}{2},n-\frac{N}{2},j} A_n^{\frac{N}{2},n-\frac{N}{2},j'} e^{i\lambda_{\frac{N}{2},n-\frac{N}{2},j'}t} e^{-i\lambda_{\frac{N}{2},n-\frac{N}{2},cj}t} \sum_{p=0}^{min[N,n]} pA_{n-p}^{\frac{N}{2},n-\frac{N}{2},j} (A^*)_{n-p}^{\frac{N}{2},n-\frac{N}{2},j'} \qquad (F.6)$$

This can also be expressed in terms of the square of the sin function as

$$S_4(\bar{n}, N, \gamma t) = -4\sum_{n=0}^{\infty}\langle n|\rho_f(0)|n\rangle \sum_{j=0}^{Min[N,n]-1}\sum_{j'=j+1}^{Min[N,n]} (A^*)_n^{\frac{N}{2},n-\frac{N}{2},j} A_n^{\frac{N}{2},n-\frac{N}{2},j'} Sin^2\left\{\frac{\left[q_{\frac{N}{2},(n-\frac{N}{2}),j} - q_{\frac{N}{2},(n-\frac{N}{2}),j'}\right]}{2}\Omega|\kappa|t\right\} \sum_{p=0}^{min[N,n]} pA_{n-p}^{\frac{N}{2},n-\frac{N}{2},j} (A^*)_{n-p}^{\frac{N}{2},n-\frac{N}{2},j'} \qquad (F.7)$$

Calculations can be performed for both resonant and non-resonant cases. The results will appear to be very similar to the all up TLM cases except when the average number of photons is less than the number of TLMs. For a single TLM, the non resonant result is

$$S_4(\bar{n}, 1, \gamma t)_{NR} = \sum_{n=1}^{\infty} \frac{n\langle n|\rho_f(0)|n\rangle}{n+\Delta} Sin^2\left(\sqrt{n+\Delta}\Omega|\kappa|t\right) \qquad (F.8)$$

This should be compared to eq. (31.a). For 2 TLMs the results for resonance is given by

$$S_4(\bar{n}, 2, \gamma t) = \langle 1|\rho_f(0)|1\rangle Sin^2(\sqrt{2}\Omega|\kappa|t) + 8\sum_{n=2}^{\infty}\langle n|\rho_f(0)|n\rangle \left\{ Sin^2\left[\sqrt{n-\frac{1}{2}}\Omega|\kappa|t\right]\frac{n(n-1)}{(2n-1)^2} \right.$$
$$\left. + \frac{n}{8(2n-1)^2}Sin^2\left[2\sqrt{n-\frac{1}{2}}\Omega|\kappa|t\right]\right\} \qquad (F.9)$$

This should be compared to eq. (30a). We see here that there will be N-1 separate terms each corresponding to distinct photon numbers up to N-1. Then the summation for similar terms begins with the lowest value of n=N. To complete this section, figures are provided for 10 TLMs. In order to compare the cases of all TLMs in the up state or all TLMs in the down state two examples, one for exactly 4 initial photons and the other for the coherent case with a mean of 4 photons, are shown in the following table.

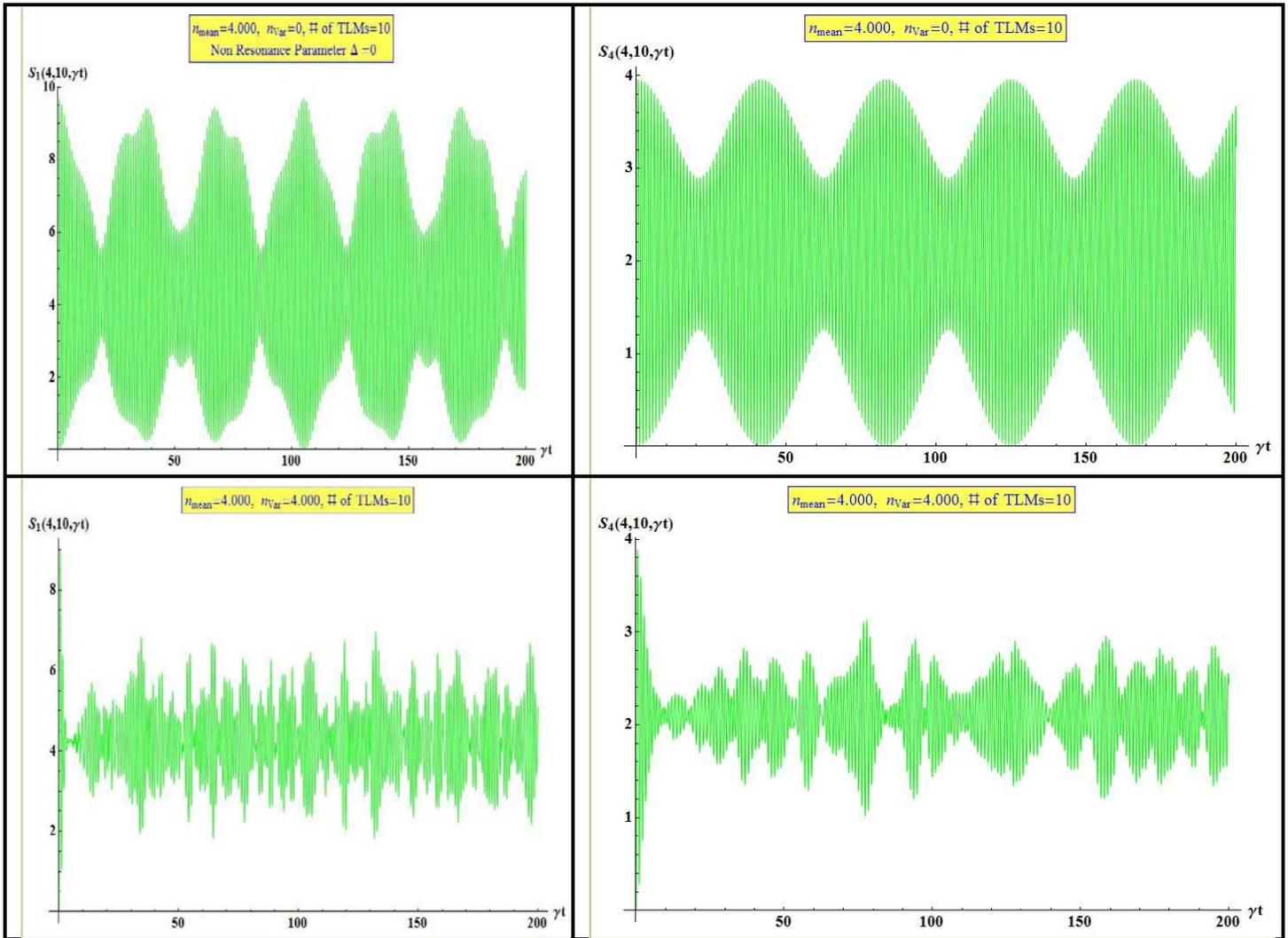

**Figure F 1: The left hand side represents all TLMs in the up state while the right hand side represents all TLMs in the down state. The top figures are for exactly 4 photons initially in the field while the lower figures represent the results of the initial field being coherent with a mean photon number of 4.**

Recall that for the all up case photons are added to the field while they are subtracted for the all down case. Further for the number of photons less than the number of TLMs, the fluctuations in $\langle E^- E^+(t) \rangle$ are more regular.